\documentclass[12pt]{spieman}  
\usepackage{amsmath,amsfonts,amssymb}
\usepackage{graphicx}
\usepackage{setspace}
\usepackage{tocloft}
\usepackage{enumitem}
\usepackage{booktabs}
\usepackage{comment}
\usepackage{listings}
\usepackage[skip=10pt]{caption} 
\usepackage{subcaption}
\usepackage{tikz} 
\usetikzlibrary{svg.path} 
\usepackage{pgfplots} 
\usepackage{pgfplotstable} 

\usepackage[newfloat, frozencache=true,cachedir=minted-cache]{minted} 

\newenvironment{code}{\captionsetup{type=listing}}{}
\SetupFloatingEnvironment{listing}{name=Code Snippet}

\usepackage{lineno}

\definecolor{mygray}{gray}{0.45}
\definecolor{mycorrect}{rgb}{0, 0, 0} 

\definecolor{codegreen}{rgb}{0,0.6,0}
\definecolor{codegray}{rgb}{0.5,0.5,0.5}
\definecolor{codepurple}{rgb}{0.58,0,0.82}
\definecolor{backcolour}{rgb}{0.95,0.95,0.92}

\newsavebox{\mybox}

\lstdefinestyle{mystyle}{
    language=Python,
    backgroundcolor=\color{backcolour},   
    commentstyle=\color{codegreen},
    keywordstyle=\color{magenta},
    numberstyle=\tiny\color{codegray},
    stringstyle=\color{codepurple},
    basicstyle=\ttfamily\footnotesize,
    breakatwhitespace=false,         
    breaklines=true,                 
    captionpos=b,                    
    keepspaces=true,                 
    numbers=left,                    
    numbersep=5pt,                  
    showspaces=false,                
    showstringspaces=false,
    showtabs=false,                  
    tabsize=2,
    columns=fullflexible
}

\title{medigan: a Python library of pretrained generative models for medical image synthesis}

\author[a,*]{Richard Osuala}
\author[a]{Grzegorz Skorupko}
\author[a]
{Noussair Lazrak}
\author[a]{Lidia Garrucho}
\author[b]{Eloy García}
\author[a]{Smriti Joshi}
\author[a]{Socayna Jouide}
\author[c]{Michael Rutherford}
\author[c]{Fred Prior}
\author[a]{Kaisar Kushibar}
\author[a]{Oliver Díaz}
\author[a]{Karim Lekadir}
\affil[a]{Barcelona Artificial Intelligence in Medicine Lab (BCN-AIM), Facultat de Matemàtiques i Informàtica, Universitat de Barcelona, Spain}
\affil[b]{Facultat de Matemàtiques i Informàtica, Universitat de Barcelona, Spain}
\affil[c]{Department of Biomedical Informatics, University of Arkansas for Medical Sciences, Little Rock, Arkansas, USA}

\cftpagenumbersoff{figure}
\cftpagenumbersoff{table} 
\begin{document} 


\maketitle

\begin{abstract}
\paragraph*{Purpose:}
Deep learning has shown great promise as the backbone of clinical decision support systems.
Synthetic data generated by generative models can enhance the performance and capabilities of data-hungry deep learning models. 
However, there is (1) limited availability of (synthetic) datasets and (2) generative models are complex to train, which hinders their adoption in research and clinical applications. To reduce this entry barrier, we explore generative model sharing to allow more researchers to access, generate, and benefit from synthetic data.

\paragraph*{Approach:}
We propose \textit{medigan}, a one-stop shop for pretrained generative models implemented as an open-source framework-agnostic Python library.
After gathering end-user requirements, design decisions based on usability, technical feasibility, and scalability 
are formulated. Subsequently, we implement \textit{medigan} based on modular components for generative model (i) execution, (ii) visualisation, (iii) search \& ranking, and (iv) contribution. We integrate pretrained models with applications across modalities such as mammography, endoscopy, x-ray, and MRI.

\paragraph*{Results:}
The scalability and design of the library is demonstrated by its growing number of integrated and readily-usable pretrained generative models, which include 21 models utilising 9 different Generative Adversarial Network architectures trained on 11 different datasets. 
We further analyse 3 \textit{medigan} applications, which include (a) enabling community-wide sharing of restricted data, (b) investigating generative model evaluation metrics, and (c) improving clinical downstream tasks. In (b), we extract Fréchet Inception Distances (FID) 
demonstrating FID variability based on image normalisation and radiology-specific feature extractors. 

\paragraph*{Conclusion:}
\textit{medigan} allows researchers and developers to create, increase, and domain-adapt their training data in just a few lines of code. Capable of enriching and accelerating the development of clinical machine learning models, we show \textit{medigan}'s viability as 
platform for 
generative model sharing. 
Our multi-model synthetic data experiments uncover new standards for assessing and reporting metrics, such as FID, in image synthesis studies.
\end{abstract}


{\noindent \footnotesize\textbf{*}Corresponding author: Richard Osuala, \linkable{Richard.Osuala@ub.edu} }

\begin{spacing}{1} 

\section{Introduction} \label{sect:intro} 

\begin{figure}
    \centering
    \includegraphics[width=\textwidth]{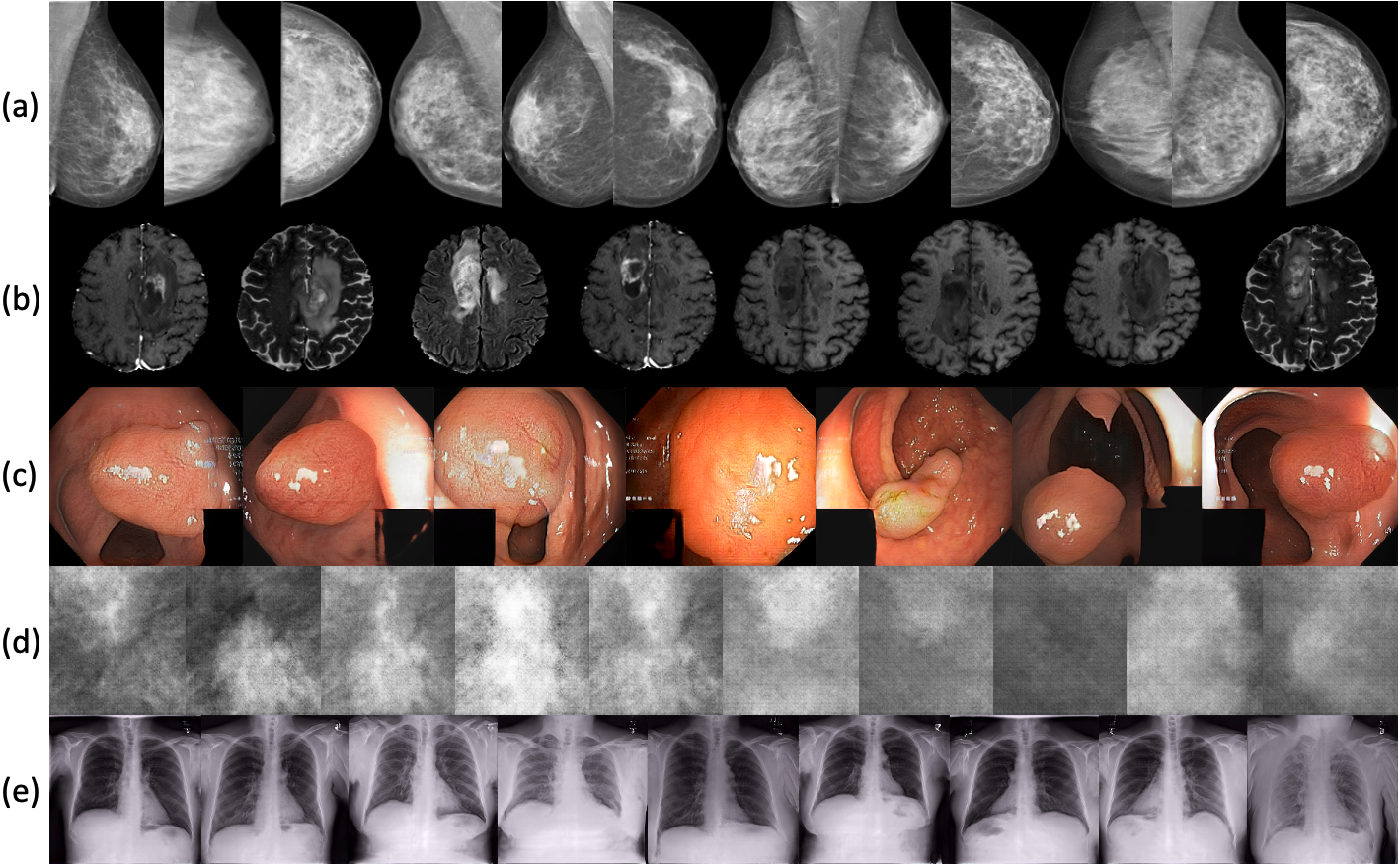}
    \caption{Randomly sampled images generated by 5 \textit{medigan} models ranging from (a) synthetic mammograms and (b) brain MRI to (c) endoscopy imaging of polyps, (d) mammogram mass patches, 
    and (e) chest x-ray imaging. The models (a)-(e) correspond to the model IDs in Table \ref{Table:model-table}, where (a): 3, (b): 7, (c): 10, (d): 12, and (e): 19.}
    \label{fig:examples}
\end{figure}

\subsection{Deep Learning and the Benefits of Synthetic Data}
The use of deep learning has increased extensively in the last decade, thanks in part to advances in computing technology (e.g., data storage, graphics processing units) and the digitisation of data. In medical imaging, deep learning algorithms have shown promising potential for clinical use due to their capability of extracting and learning meaningful patterns from imaging data and their high performance on clinically-relevant tasks. These include image-based disease diagnosis \cite{martin2020image, aggarwal2021diagnostic} and detection \cite{liu2019comparison}, as well as medical image reconstruction \cite{schlemper2017deep, ahishakiye2021survey}, segmentation \cite{tajbakhsh2020embracing}, and image-based treatment planning \cite{osuala2022data, jin2021predicting, bi2019artificial}.


However, deep learning models need vast amounts of well-annotated data to reliably learn to perform clinical tasks, while, at the same time, the availability of public medical imaging datasets remains limited due to legal, ethical, and technical patient data sharing constraints \cite{bi2019artificial, prior2020open}. In the common scenario of limited imaging data, synthetic images, such as the ones illustrated in Figure \ref{fig:examples}, are a useful tool to improve the learning of the artificial intelligence (AI) algorithm e.g. by enlarging its training dataset \cite{osuala2022data, yi2019generative, alyafi2020dcgans}. Furthermore, synthetic data can be used to minimise problems associated with domain shift, data scarcity, class imbalance, and data privacy \cite{osuala2022data}.
For instance, a dataset can be balanced by populating the less frequent classes with synthetic data during training (class imbalance). Further, as domain-adaptation technique, a dataset can be translated from one domain to another, e.g., from MRI to CT \cite{wolterink2017deep}  (domain shift). Regarding data privacy, synthetic data can be shared instead of real patient data to improve privacy preservation \cite{stadler2022synthetic, szafranowska2022sharing, osuala2022data}.

\subsection{The Need of Reusable Synthetic Data Generators}
Commonly, generative models are used to produce synthetic imaging data, with Generative Adversarial Networks (GANs) \cite{goodfellow2014generative} 
being popular models of choice.
However, the adversarial training scheme required by GANs and related networks is known to pose challenges in regard to (i) achieving training stability, (ii) avoiding mode collapse, and (iii) reaching convergence \cite{salimans2016improved, mescheder2018training, arora2018gans}.

Hence, the training process of GANs and generative models at large is  non-trivial and requires a considerable time investment for each training iteration as well as specific hardware and a fair amount of knowledge and skills in the area of AI and generative modelling. Given these constraints, researchers and engineers often refrain from generating and integrating synthetic data into their AI training pipelines and experiments. This issue is further exacerbated by the prevailing need of training a new generative model for each new data distribution, 
which, in practice, often means that a new generative model has to be trained for each new application, use-case, and dataset.

\subsection{Community-Driven Model Sharing and Reuse}
We argue that a feasible solution to this problem is the community-wide sharing and reuse of pretrained generative models. Once successfully trained, such a model can be of value to multiple researchers and engineers with similar needs. For example, researchers can reuse the same model if they work on the same problem, conduct similar experiments, or evaluate their methods on the same dataset. 
\textcolor{mycorrect}{We note that such reuse ideally is subject to previous inspection of generative model limitations with the model's output quality having qualified as suitable for the task at hand. The quality of a model's output data and annotations can commonly be measured via (a) expert assessment, (b) computation of image quality metrics, or (c) downstream task evaluation.
In sum,} the problem of synthetic data generation calls for a community-driven solution, where a generative model trained by one member of the community can be reused by other members of the community. Motivated by the absence of such a community-driven solution for synthetic medical data generation, we designed and developed \textit{medigan} to bridge the gap between the need for synthetic data and complex generative model creation and training processes.

\section{Background and Related Work}

\subsection{Generative Models}

While discriminative models are able to distinguish between data instances of different kinds (label samples), generative models are able to generate new data instances (draw samples). In contrast to modelling decision boundaries in a data space, generative models model how data is distributed within that space.
Deep generative models \cite{ruthotto2021introduction} are composed of multi-hidden layer neural networks to explicitly or implicitly estimate a probability density function (PDF) from a set of real data samples. After approximating the PDF from observed data points (i.e., learning the real data distribution), these models can then sample unobserved new data points from that distribution. In computer vision and medical imaging, synthetic images are generated by sampling such unobserved points from high-dimensional imaging data distributions. Popular deep generative models to create synthetic images in these fields include Variational Autoencoders \cite{kingma2013auto}, Normalizing Flows \cite{rezende2015variational, dinh2014nice, dinh2016density}, Diffusion Models\cite{sohl2015deep, song2019generative, ho2020denoising}, and Generative Adversarial Networks (GANs) \cite{goodfellow2014generative}. From these, the versatile GAN framework has seen the most widespread adoption in medical imaging to date \cite{osuala2022data}. We, hence, centre our attention on GANs in the remainder of this work, but emphasise that contributions of other types of generative models are equally welcome in the \textit{medigan} library.

\begin{figure*}
	\centering
       		{\includegraphics[width=1.0\textwidth]{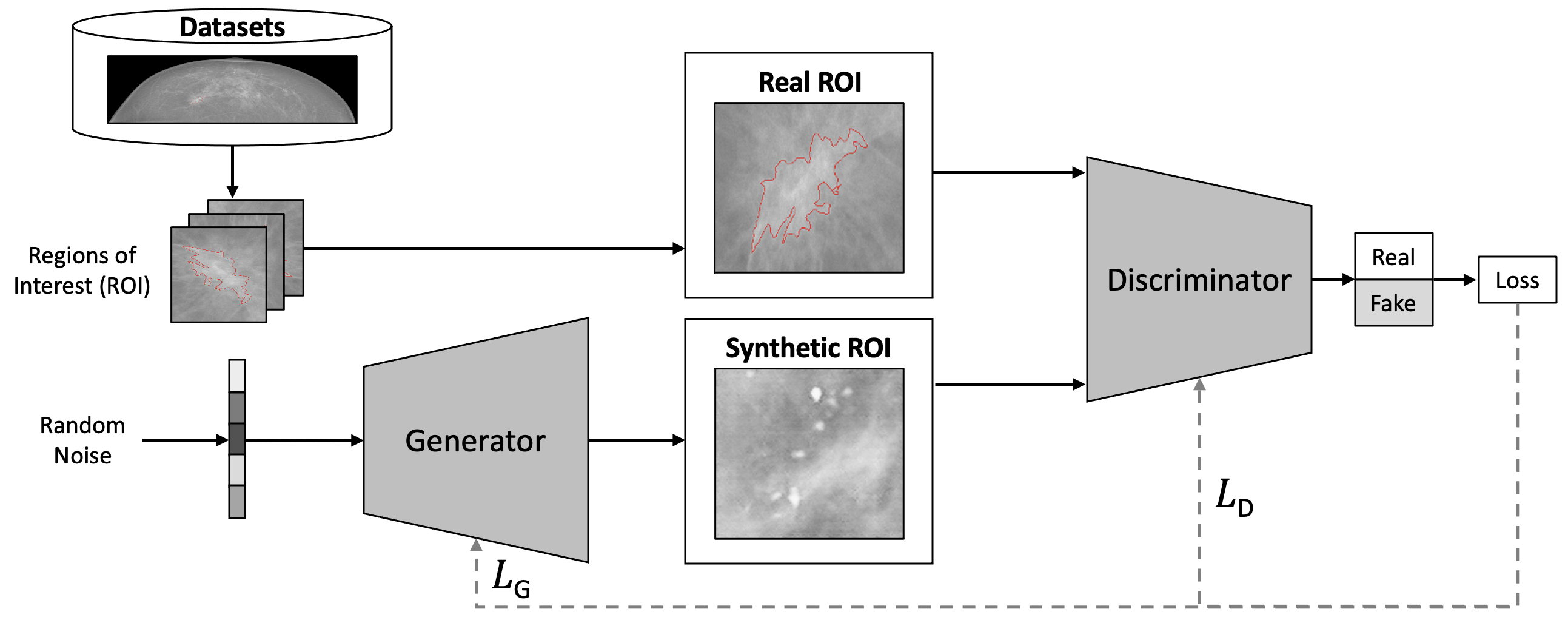}}
	 \caption[]{The GAN framework. In this visual example, the generator network receives random noise vectors, which it learns to map to region-of-interest patches of full-field digital mammograms. During training, the adversarial loss is not only backpropagated to the discriminator as $L_{D}$, but also to the generator as $L_{G}$.
	 This particular architecture and loss function was used to train \textit{medigan} models listed with IDs 1, 2, and 5 in Table \ref{Table:model-table}.}
  	\label{fig:gan}
\end{figure*}

\subsection{Generative Adversarial Networks}
The training of GANs comprises two neural networks, the generator network (G) and the discriminator network (D), as illustrated by Figure \ref{fig:gan} for the example of mammography region-of-interest patch generation. G and D compete against each other in a two-player zero-sum game defined by the value function shown in equation \ref{eq:1}. Subsequent studies extended the adversarial learning scheme by proposing innovations of the loss function, G and D network architectures, and GAN applications by introducing conditions into the image generation process.

\begin{equation} \label{eq:1}
\begin{aligned}
\min_{G} \max_{D} V(D,G) = \min_{G} \max_{D}[\mathbb{E}_{x\sim p_{data}} [log D(x)] + \mathbb{E}_{z\sim p_{z}} [log(1 - D(G(z)))]]
\end{aligned}
\end{equation}

\subsubsection{GAN Loss Functions}
Goodfellow et al (2014)~\cite{goodfellow2014generative} define the discriminator as a binary classifier classifying whether a sample $x$ is either real or generated. The discriminator is, hence, trained via binary-cross entropy with the objective of minimising the adversarial loss function 
shown in equation \ref{eq:2}, which the generator, on the other hand, tries to maximise. In Wasserstein GAN (WGAN) \cite{arjovsky2017wasserstein} the adversarial loss function is replaced with a loss function based on the Wasserstein-1 distance between real and fake sample distributions estimated by D (alias 'critic'). Gulrajani et al (2017) \cite{gulrajani2017improved} resolve the need to enforce a 1-Lipschitz constraint in WGAN via gradient penalty (WGAN-GP) instead of WGAN weight clipping. 
Equation \ref{eq:3} depicts the WGAN-GP discriminator loss 
with penalty coefficient $\lambda$ and distribution $\mathbb{P}_{\hat x}$ based on 
sampled pairs from (a) the real data distribution $\mathbb{P}_{data}$ and (b) the generated data distribution $\mathbb{P}_{g}$. 

\begin{equation} \label{eq:2}
\begin{aligned}
L_{D_{GAN}} = - \mathbb{E}_{x\sim p_{data}} [log D(x)] + \mathbb{E}_{z\sim p_{z}} [log(1 - D(G(z)))]
\end{aligned}
\end{equation}
\begin{eqnarray} \label{eq:3}
\begin{array}{l}
 L_{D_{WGAN-GP}} \ = \ {E_{\tilde x \sim {\mathbb{P}_g}}}\left[ {D\left( {\tilde x} \right)} \right] \ - \ {E_{x \sim {\mathbb{P}_{data}}}}\left[ {D\left( x \right)} \right] \ + \ \lambda \ {E_{\hat x \sim {\mathbb{P}_{\hat x}}}}\left[ {{{\left( {{{\left\| {{\nabla _{\hat x}}D\left( {\hat x} \right)} \right\|}_2} - 1} \right)}^2}} \right] \\ 
 \end{array}
 \end{eqnarray}

In addition to changes to the adversarial loss, further studies integrate additional loss terms into the GAN framework. For instance, FastGAN \cite{liu2020towards} uses an additional reconstruction loss in the discriminator, which, for improved regularisation, is trained as self-supervised feature-encoder. 

\subsubsection{GAN Network Architectures and Conditions}
A plethora of different GAN network architectures has been proposed \cite{osuala2022data, kang2022studiogan} starting with a deep convolutional GAN (DCGAN)\cite{radford2015unsupervised} neural network architecture of both D and G. Later approaches, for example, include a ResNet-based architecture as backbone \cite{gulrajani2017improved} and progressively-grow the generator and discriminator networks during training to enable high resolution image synthesis (PGGAN) \cite{karras2017progressive}.

Another line of research has been focusing on conditioning the output of GANs based on discrete or continuous labels. For example, in cGAN this is achieved by feeding a label to both D and G \cite{mirza2014conditional}, while in the auxiliary classifier GAN (AC-GAN) the discriminator additionally predicts the label that is provided to the generator \cite{odena2017conditional}.

Other models condition the generation process on input images \cite{isola2017image, zhu2017unpaired, choi2018stargan, park2019semantic, sushko2020you} unlocking image-to-image translation and domain-adaptation GAN applications. A key difference in image-to-image translation methodology is the presence (paired translation) or absence (unpaired translation) of corresponding image pairs in the target and source domain. Using a L1 reconstruction loss between target and source domain alongside the adversarial loss from equation \ref{eq:2}, pix2pix \cite{isola2017image} defines a common baseline model for paired image-to-image translation. For unpaired translation, cycleGAN \cite{zhu2017unpaired} is a popular approach, which also consists of a L1 reconstruction (cycle-consistency) loss  between a source (target) image and a source (target) image translated to target (source) and back to source (target) via two consecutive generators.

A further methodological innovation includes SinGAN \cite{shaham2019singan}, which, based on only a single training image, learns to generate multiple synthetic images. This is accomplished via a multi-scale coarse-to-fine pipeline of generators, where a sample is passed sequentially through all generators, each of which also receives a random noise vector as input.

\subsection{Generative Model Evaluation} \label{sec:evaluation}
One approach of evaluating generative models is by human expert assessment of their generated synthetic data. In medical imaging, such observer studies often enlist board-certified clinical experts such as radiologists or pathologists to examine the quality and/or realism of the synthetic medical images \cite{korkinof2020perceived, alyafi20quality}.
However, this approach is manual, laborious and costly, and, hence, research attention has been devoted to automating generative model evaluation \cite{borji2019pros, borji2022pros}, including:
\begin{enumerate}[label=(\roman*)]
    \item Metrics for automated analysis of the synthetic data and its distribution, such as the Inception Score (IS) \cite{salimans2016improved} and Fréchet Inception Distance (FID) \cite{heusel2017gans}. Both metrics are popular in computer vision \cite{kang2022studiogan}, while the latter also has seen widespread adoption in medical imaging \cite{osuala2022data}.
    
    FID is based on a pretrained Inception \cite{szegedy2016rethinking} model (e.g., v1 \cite{szegedy2015going}, v3 \cite{szegedy2016rethinking}) 
    to extract features from synthetic and real datasets, which are then fitted to multi-variate Gaussians $X$ (e.g., real) and $Y$ (e.g., synthetic) with means $\mu_{X}$ and $\mu_{Y}$ and covariance matrices $\Sigma_{X}$ and $\Sigma_{Y}$. Next, $X$ and $Y$ are compared via the Wasserstein-2 (Fréchet) distance (FD), as depicted by equation \ref{eq:4}.
    \begin{equation} \label{eq:4}
    \begin{aligned}
        \textit{FD}(X, Y) = \lVert \mu_{X} - \mu_{Y} \rVert_{2}^{2} + \text{tr}(\Sigma_{X} + \Sigma_{Y} -2(\Sigma_{X}\Sigma_{Y})^{\frac{1}{2}})
    \end{aligned}
    \end{equation}
    
    \item Metrics that compare a synthetic image with a real reference image such as: mean squared error (MSE), peak signal-to-noise ratio (PSNR), and structural similarity index measure (SSIM) \cite{wang2004image}. Given the absence of corresponding reference images, such metrics are not readily applicable for unconditional noise-to-image generation models.
    
    \item Metrics that compare the performance of a model on a surrogate downstream task with and without generative model intervention \cite{szafranowska2022sharing, osuala2022data, thambawita2022singan, garrucho2022high}. For instance, training on additional synthetic data can increase a model's downstream task performance, thus, demonstrating the usefulness of the generative model that generated such data.
\end{enumerate}

For the analysis of generative models in the present study we discard (ii) due to its limitation of requiring specific reference images. We further deprioritise the IS from (i) due to its limited applicability to medical imagery stemming from it missing a comparison between real and synthetic data distributions combined with it having a strong bias on natural images via its ImageNet\cite{deng2009imagenet}-pretrained Inception classifier as backbone feature extractor. Therefore, we focus on FID from (i) and downstream task performance (iii) as potential evaluation measures for medical image synthesis models in the remainder of this work.

\subsection{Image Synthesis Tools and Libraries}

Related libraries, such as pygan \cite{pygan}, torchGAN \cite{Pal2021}, vegans \cite{vegans}, imaginaire \cite{imaginaire}, TF-GAN \cite{tfgan}, PyTorch-GAN \cite{pytorchgan}, keras-GAN \cite{kerasgan}, mimicry \cite{lee2020mimicry}, and studioGAN \cite{kang2022studiogan} have focused on facilitating the implementation, training, and comparative evaluation of GANs in computer vision (CV). Despite a strong focus on language models, the HuggingFace transformers library and model hub \cite{wolf2019huggingface} also contain a few pretrained computer vision GAN models. The GAN Lab\cite{kahng2018gan} provides an interactive visual experimentation tool 
to examine the training process and its data flows in GANs.


Specific to AI in medical imaging, Diaz et al (2021)\cite{diaz2021data} provided a comprehensive survey of tools, libraries and platforms for privacy preservation, data curation, medical image storage, annotation, and repositories.
Compared to CV, fewer GAN and AI libraries and tools exist in medical imaging. Furthermore, CV libraries are not always suited to address the unique challenges of medical imaging data \cite{perez2021torchio, diaz2021data, moore2022cleanx}. For instance, pretrained generative models from computer vision cannot be readily adapted to produce medical imaging specific outputs. The TorchIO library \cite{perez2021torchio} addresses the gap between CV and medical image data processing requirements providing functions for efficient loading, augmentation, preprocessing, and patch-based sampling of medical imagery. The Medical Open Network for AI (MONAI) \cite{monai_consortium_2020_5525502} is a PyTorch-based\cite{pytorchPaszke} framework that facilitates the development of diagnostic AI models with tutorials for classification, segmentation, and AI model deployment. Further efforts in this realm include NiftyNet \cite{gibson2018niftynet}, the deep learning tool kit (DLTK)\cite{pawlowski2017state}, MedicalZooPytorch \cite{adaloglou2019MRIsegmentation}, and nnDetection \cite{baumgartner2021nndetection}. The recent RadImageNet initiative\cite{mei2022radimagenet} shares baseline image classification models pretrained on a dataset designed as the radiology medical imaging equivalent to ImageNet\cite{deng2009imagenet}. 
 
To the best of our knowledge, no open-access software, tool, or library exists that targets reuse and sharing of pretrained generative models in medical imaging. To this end, we expect the contribution of our \textit{medigan} library to be instrumental in enabling dissemination of generative models and increased adoption of synthetic data into AI training pipelines. As an open-access plug-and-play solution for generation of multi-purpose synthetic data, \textit{medigan} aims to benefit patients and clinicians by enhancing the performance and robustness of AI-based clinical decision support systems.

\section{Method: The \textit{medigan} Library}

\begin{table}[H]
\scalebox{0.95}{
\begin{tabular}{|c|l|l|}
\hline
\textbf{ } & \textbf{Title} & \textbf{\textit{medigan} metadata} \\
\hline
\hline
1 & Code version & v1.0.0 \\
\hline
2 & Code license & \href{https://github.com/RichardObi/medigan/blob/main/LICENSE}{MIT}  \\
\hline
3 & Code version control system & git \\
\hline
4 & Software languages & Python \\
\hline
5 & Code repository & \href{https://github.com/RichardObi/medigan}{https://github.com/RichardObi/medigan} \\
\hline
6 & Software package repository & \href{https://pypi.org/project/medigan/} {https://pypi.org/project/medigan/} \\
\hline
7 & Developer documentation & \href{https://medigan.readthedocs.io} {https://medigan.readthedocs.io} \\

\hline
8 & Tutorial & medigan quickstart (\href{https://github.com/RichardObi/medigan/blob/main/examples/tutorial.ipynb}{tutorial.ipynb}) \\
\hline
9 & Requirements for compilation & Python v3.6+ \\
\hline
10 & Operating system & OS independent. Tested on Linux, OSX, Windows.\\
\hline
11 & Support email address & Richard.Osuala[at]gmail.com \\
\hline
12 & Dependencies & tqdm, requests, torch, numpy, PyGithub, matplotlib (\href{https://github.com/RichardObi/medigan/blob/main/setup.py\#L26} {setup.py})\\
\hline
\end{tabular}
}
\caption{Overview of \textit{medigan} library information.}
\label{Table:metadata} 
\end{table}

\begin{figure}
	\centering
       	\includegraphics[width=1.0\textwidth]{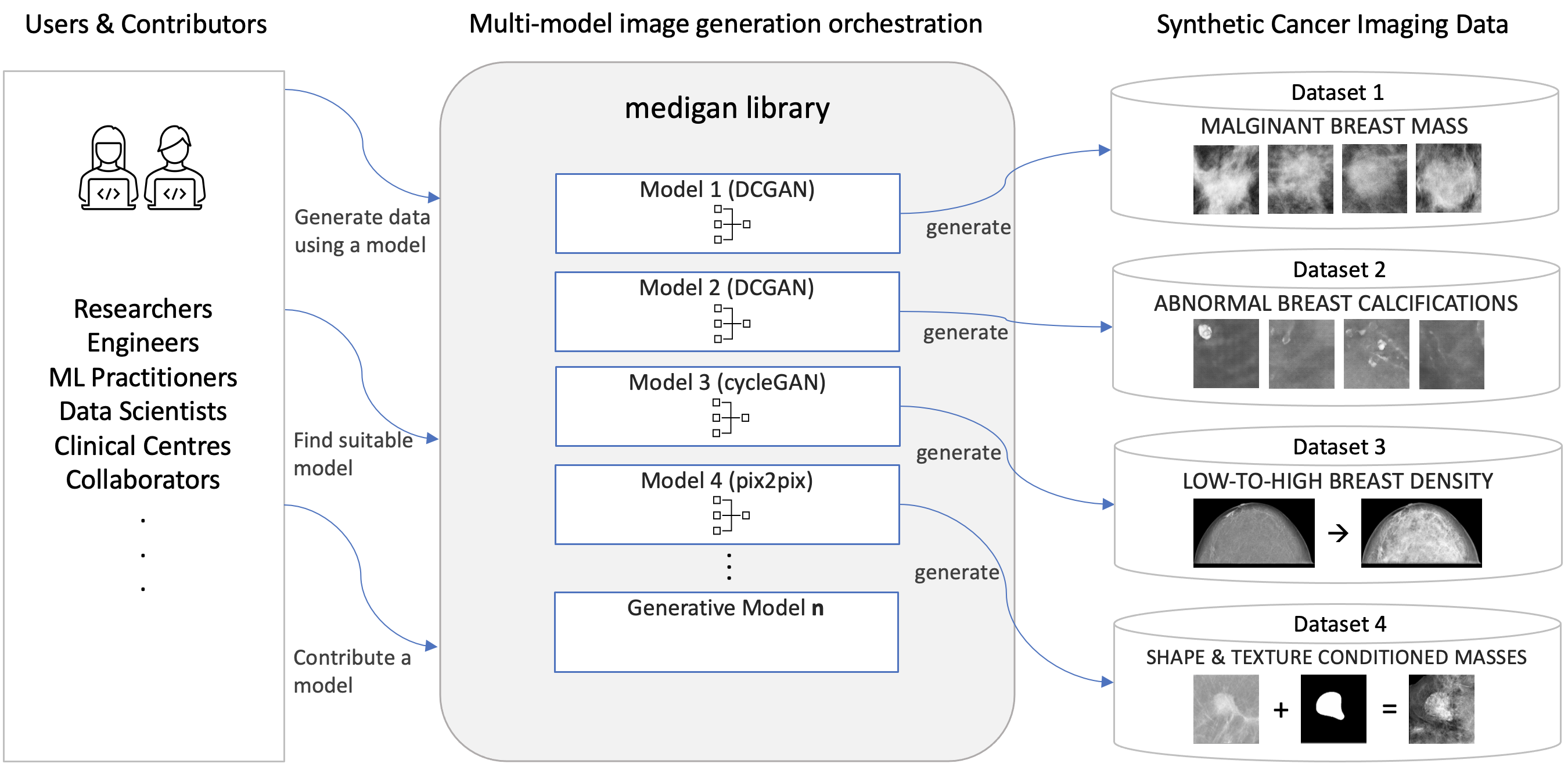}
	 \caption{Architectural overview of \textit{medigan}. Users interact with the library by contributing, searching, and executing generative models, the latter shown here exemplified for mammography image generation with models with ids 1 to 4 described in Table \ref{Table:model-table}.
	 }
  	\label{fig:overview}
\end{figure}

We contribute \textit{medigan} as an open-source open-access MIT-licensed Python3 library distributed via the Python Package Index (Pypi) for synthetic medical dataset generation, e.g., via pretrained generative models. The metadata of \textit{medigan} is summarised in Table \ref{Table:metadata}. \textit{medigan} accelerates research in medical imaging by flexibly providing (a) synthetic data augmentation and (b) preprocessing functionality, both readily integrable in machine learning training pipelines. It also allows contributors to add their generative models in a thought-through process and provides simplistic functions for end-users to search for, rank, and visualise models. The overview of \textit{medigan} in Figure \ref{fig:overview} depicts the core functions demonstrating how end-users can 
(a) contribute a generative model, (b) find a suitable generative model inside the library, and (c) generate synthetic data with that model.

\subsection{User Requirements and Design Decisions}
End-user requirement gathering is recommended for the development of trustworthy AI solutions in medical imaging \cite{lekadir2021futureai}. Therefore, we organised requirement gathering sessions with potential end-users, model contributors and stakeholders from the EuCanImage Consortium, a large European H2020 project (\href{https://eucanimage.eu/}{https://eucanimage.eu/}) 
building a cancer imaging platform for enhanced Artificial Intelligence in oncology. Upon exploring the needs and preferences of medical imaging researchers and AI developers, respective requirements for the design of \textit{medigan} were formulated to ensure usability and usefulness. For instance, the users articulated a clear preference for a user interface in the format of an importable package as opposed to a graphical user interface (GUI), web application, database system, or API. Table \ref{Table:requirements-table} summarises key requirements and the corresponding design decisions. 

\begin{table*}[h]
\centering
\scriptsize
\caption{Overview of the key requirements gathered together with potential end-user alongside the respective design decisions taken towards fulfilling these requirements with \textit{medigan}. }\label{Table:requirements-table}
\scalebox{1.0}{
\begin{tabular}{{p{0.01\textwidth}p{0.40\textwidth}p{0.50\textwidth}}}
    \hline
    \textbf{No} &
    \textbf{End-User Requirement} &
    \textbf{Respective Design Decision}
    \\
    \hline
    1 &
    Instead of a GUI tool, \textit{medigan} should be implemented as a platform-independent library importable into users' code. &
    Implementation of \textit{medigan} as publicly accessible Python package distributed via PyPI.
    \\
    \hline
    2 &
    It should support common frameworks for building generative models, e.g., PyTorch \cite{pytorchPaszke}, TensorFlow \cite{tensorflow2015-whitepaper}, Keras \cite{chollet2015keras}. &
    \textit{medigan} is built framework-agnostic treating each  model as separate Python package with freedom of choice of framework and dependencies.
    \\
    \hline
    3 &
    The library should allow different types of generative models and generation processes. &  \textit{medigan} supports any type of data generation model including GANs\cite{goodfellow2014generative}, VAEs\cite{kingma2013auto}, flow-based\cite{rezende2015variational, dinh2014nice, dinh2016density}, diffusion\cite{sohl2015deep, song2019generative, ho2020denoising} and non-deep learning models.
    \\
    \hline
    4 &
    The library should support different types of synthetic data. &
    \textit{medigan} supports any type of synthetic data ranging from 2D and 3D images to image pairs, masks, and tabular data.
    \\
    \hline
    5 &
    Sample generation functions should be easily integrable into diverse user code, pipelines and workflows. &
    \textit{medigan}'s generate function can (i) return samples, (ii) generate folders with samples, or (iii) return a model's generate function as callable.
    \\
    \hline
    6 &
    User should be able to integrate \textit{medigan} data in AI training via a dataloader. &
    For each model, \textit{medigan} supports returning a \href{https://pytorch.org/docs/stable/data.html\#torch.utils.data.DataLoader}{torch} dataloader readily integrable in AI training pipelines, combinable with other dataloaders.
    \\
    \hline
    7 & 
    Despite using large deep learning models, the library should be as lightweight as possible. &
    Only the user-requested models are downloaded and locally imported. Thus, model dependencies are not part of \textit{medigan}'s dependencies.
    \\
    \hline
    8 &
    It should be possible to locally review and adjust a generative model of the library. &
    After download, a model's code and config is available for end-users to explore and adjust. \textit{medigan} can also load models from local file systems.
    \\
    \hline
    9 &
    The library should support both CPU and GPU usage depending on a user's hardware. &
    Contributed \textit{medigan} models are reviewed and, if need be, enhanced to run on both GPU and CPU.
    \\
    \hline
    10 &
    Version and source of the models that the library load should be transparent to the end-user. &
    Convention of storing \textit{medigan} models on \href{https://zenodo.org/}{Zenodo}, where each model's source code and version history is available.
    \\
    \hline
    11 &
    There should be no need to update the version of the \textit{medigan} package each time a new model is contributed. &
    \textit{medigan} is designed independently of its model packages separately stored on Zenodo. \href{https://github.com/RichardObi/medigan/blob/main/config/global.json}{Config} updates do not require new \textit{medigan} versions.   
    \\
    \hline
    12 &
    Following \cite{lekadir2021futureai}, models are contributed in transparently and tracebly, allowing quality and reproducibility checks. &
    Model contribution is traceable via version control. Adding models to \textit{medigan} requires a \href{https://github.com/RichardObi/medigan/blob/main/config/global.json}{config} change via pull request.
    \\
    \hline
    13 &
    The risk that the library downloads models that contain malicious code should be minimised. &
    Zenodo model uploads receive static DOIs. After verification, unsolicited uploads/changes do not affect \textit{medigan}, which \href{https://github.com/RichardObi/medigan/blob/main/config/global.json}{points to} specific DOI. 
    \\
    \hline
    13 &
    License and authorship of generative model contributors should be clearly stated and acknowledged. &
    Separation of models and library allows freedom of choice of model license and transparent authorship reported for each model.
    \\
    \hline
    14 &
    Each generative model in the library should be documented. &
    Each available model is listed and described in \textit{medigan}'s \href{https://medigan.readthedocs.io/en/latest/models.html}{documentation}, in the \href{https://github.com/RichardObi/medigan/blob/main/README.md}{readme}, and also separately in its Zenodo entry.
    \\
    \hline
    15 &
    The library should have minimal dependencies on the user side and should run on common end-user systems. & \textit{medigan} has a minimal set of Python dependencies, is OS-independent, and avoids system and third-party dependencies.  
    \\
    \hline
    16 &
    Contributing models should be simple and at least partially automated. & \textit{medigan}'s contribution workflow automates local model configuration, testing, packaging, Zenodo upload, and issue creation on GitHub.
    \\
    \hline
    17 &
    If different models have the same dependency but with different versions, this should not cause a conflict. & Model dependency versions are specified in the \href{https://github.com/RichardObi/medigan/blob/main/config/global.json}{config}. 
    \textit{medigan}'s generate method can install unsatisfied dependencies, 
    avoiding conflicts.
    \\
    \hline
    19 &
    Any model in the library should be automatically tested and results reported to make sure all models work as designed. & On each commit to \href{https://github.com/RichardObi/medigan/tree/main}{main}, a \href{https://github.com/RichardObi/medigan/actions}{CI pipeline} automatically builds, formats, and lints \textit{medigan} before \href{https://github.com/RichardObi/medigan/blob/main/tests/test_generator.py}{testing} all models and core functions.
    \\
    \hline
    20 &
    The library should make the results of the models visible with minimal code required by end-users. & \textit{medigan}'s simple visualisation feature allows users to adjust a model's input latent vector for intuitive exploration of output diversity and fidelity.
    \\
    \hline
    21 &
    The library should support large synthetic dataset generation on user machines with limited random-access memory. & For large synthetic dataset generation, \textit{medigan} iteratively generates samples via small batches to avoid exceeding users' in-memory storage limits.
    \\
    \hline
    22 &
    Users can specify model weights, model inputs, number and storage location of the synthetic samples. & Diverging from \href{https://github.com/RichardObi/medigan/blob/main/config/global.json}{defaults}, users can specify (i) weights, (ii) number of samples (iii) return or store, (iv) store location, (v) optional inputs.
    \\
    \hline
    \end{tabular}
}
\end{table*}

\subsection{Software Design and Architecture}
\textit{medigan} is built with a focus on simplicity and usability. The integration of pretrained models is designed as internal Python package import and offers simultaneously (a) high flexibility to and (b) low code dependency on these generative models. The latter allows the reuse of the same orchestration functions in \textit{medigan} for all model packages. 

\textcolor{mycorrect}{Using object-oriented programming}, the same \mintinline[fontsize=\footnotesize]{python}{model_executor} class is used to implement, instantiate, and run all different types of generative model packages. To keep the library maintainable and lightweight, and to avoid limiting interdependencies between library code and generative model code, \textit{medigan}'s models are hosted outside the library (on Zenodo) as independent Python modules.
To avoid long initialisation times upon library import, lazy loading is applied. A model is only loaded and its \mintinline[fontsize=\footnotesize]{python}{model_executor} instance is only initialised if a user specifically requests synthetic data generation for that model.
To achieve high cohesion\cite{larman2001applying} i.e. keeping the library and its functions specific, manageable and understandable, the library is structured into several modular components. 
These include the loosely-coupled \mintinline[fontsize=\footnotesize]{python}{model_executor}, \mintinline[fontsize=\footnotesize]{python}{model_selector}, and \mintinline[fontsize=\footnotesize]{python}{model_contributor} modules.

The \mintinline[fontsize=\footnotesize]{python}{generators} module is inspired by the facade design pattern\cite{gamma1995design} and acts as a single point of access to all of \textit{medigan}'s functionalities. As single interface layer between users and library, it reduces interaction complexity and provides users with a clear set of readily extendable library functions. Also, the \mintinline[fontsize=\footnotesize]{python}{generators} module increases internal code reusability and allows for combination of functions from other modules. For instance, a single function call can run the generation of samples by the model with the highest FID score of all models found in a keyword search.

\subsection{Model Metadata} \label{sec:metadata}
The FID score and all other model information such as dependencies, modality, type, zenodo link, associated publications, and generate function parameters are stored in a single comprehensive model metadata json \textcolor{mycorrect}{file. Alongside its searchability, readability, and flexibility, the choice of json as file format is motivated by its extendability to a non-relational database.}
As single source of model information, the \textit{global.json} file consists of an array of model IDs, where under each model id the respective model metadata is stored. 
Towards ensuring model traceability as recommended by the FUTURE-AI consensus guidelines \cite{lekadir2021futureai}, each model (on Zenodo) and its global.json metadata (on GitHub) are version-controlled with the latter being structured into the following objects.
    \begin{enumerate}[label=(\roman*)]
        \item \textit{execution}: contains the information needed to download, package and run the model resources.
        \item \textit{selection}: contains model evaluation metrics and further information used to search, compare, and rank models.
        \item \textit{description}: contains general information and main details about the model such as title, training dataset, license, date, and related publications.
    \end{enumerate}
This \textit{global.json} metadata file is retrieved
, provided, and handled by the \mintinline[fontsize=\footnotesize]{python}{config_manager} module once a user imports the \mintinline[fontsize=\footnotesize]{python}{generators} module. This facilitates rapid access to a model's metadata given its \mintinline[fontsize=\footnotesize]{python}{model_id} and allows to add new models or model versions to \textit{medigan} via pull request without requiring a new release of the library.

\subsection{Model Search and Ranking} \label{sec:search_workflow}

\begin{figure}
	\centering
	  \scalebox{0.90}{
       	\includegraphics[width=1.0\textwidth]{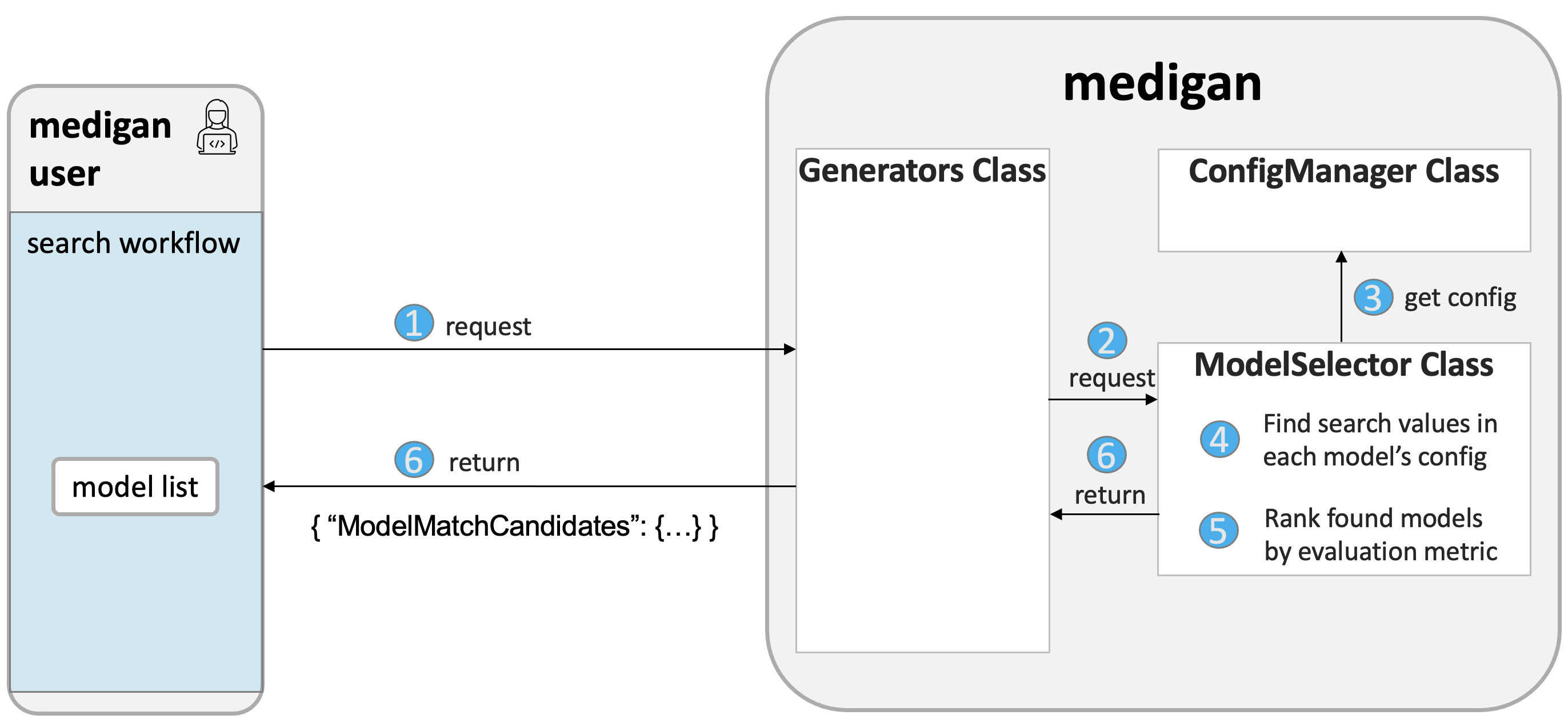}}
	 \caption{The search workflow. A user sends a search query (1) to the \mintinline[fontsize=\footnotesize]{python}{Generators} class, which triggers a search (2) via the \mintinline[fontsize=\footnotesize]{python}{ModelSelector} class. The latter retrieves the \textit{global.json} model metadata/config dict (3), in which it searches for query values finding matching models (4). Next, the matched models are optionally also ranked based on a user-defined performance indicator (5) before being returned as list to the user.}
  	\label{fig:search_workflow}
\end{figure}

The number of models in \textit{medigan} is expected to grow over time. Potentially this will lead to the foreseeable issue where users of \textit{medigan} have a large number of models to choose from. Users likely will be uncertain which model best fits their needs depending on their data, modality, use-case and research problem at hand and would have to to go through each models metadata to find the most suitable model in \textit{medigan}. Hence, to facilitate model selection, the \mintinline[fontsize=\footnotesize]{python}{model_selector} module implements model search and ranking functionalities. This search workflow is shown in Figure \ref{fig:search_workflow} and triggered by running Code Snippet \ref{code:search}.

The \mintinline[fontsize=\footnotesize]{python}{model_selector} module contains a search method that takes search operator (i.e OR, AND, or XOR) and a keyword search values list as parameters and recursively searches through the models' metadata. The latter is provided by the \mintinline[fontsize=\footnotesize]{python}{config_manager} module. The \mintinline[fontsize=\footnotesize]{python}{model_selector} populates a \mintinline[fontsize=\footnotesize]{python}{modelMatchCandidates} object with \mintinline[fontsize=\footnotesize]{python}{matchedEntry} instances each of which represents a potential model match to the search query. The \mintinline[fontsize=\footnotesize]{python}{modelMatchCandidates} class evaluates which of it's associated model matches should be flagged as true match given the search values and search operator.
The method \mintinline[fontsize=\footnotesize]{python}{rank_models_by_performance} compares either all or specified models in \mintinline[fontsize=\footnotesize]{python}{medigan} by a performance indicator such as FID. This indicator commonly is a metric that correlates with diversity, fidelity, or condition adherence to estimate the quality of generative models and/or the data they generate \cite{osuala2022data}. The \mintinline[fontsize=\footnotesize]{python}{model_selector} looks up the value for the specified performance indicator in the model metadata and returns a descendingly or ascendingly ranked list of models to the user.

\begin{code}
\captionof{listing}{Searching for a model in \textit{medigan}.}
\vspace*{-5mm}
\label{code:search}
\begin{minted}[mathescape,
               linenos,
               numbersep=5pt,
               gobble=0,
               frame=lines, 
               fontsize=\footnotesize,
               framesep=2mm]{python}
from medigan import Generators # import
generators = Generators() # init
values=['patches', 'mammography'] # keywords of search query
operator='AND' # all keywords are needed for match
results = generators.find_model(values, operator)
\end{minted}
\end{code}

\subsection{Synthetic Data Generation}

Synthetic data generation is \textit{medigan}'s core functionality towards overcoming scarcity of (a) training data and (b) reusable generative model in medical imaging. Posing a low entry barrier for non-expert users, \textit{medigan}'s \mintinline[fontsize=\footnotesize]{python}{generate} method is both simple and scalable. While a user can run it with only one line of code, it flexibly supports any type of generative model and synthetic data generation process, as illustrated in Table \ref{Table:model-table} and by Figure \ref{fig:examples}.

\subsubsection{The Generate Workflow} \label{sec:generate}

\begin{figure}
	\centering
       	 \scalebox{0.95}{\includegraphics[width=1.0\textwidth]{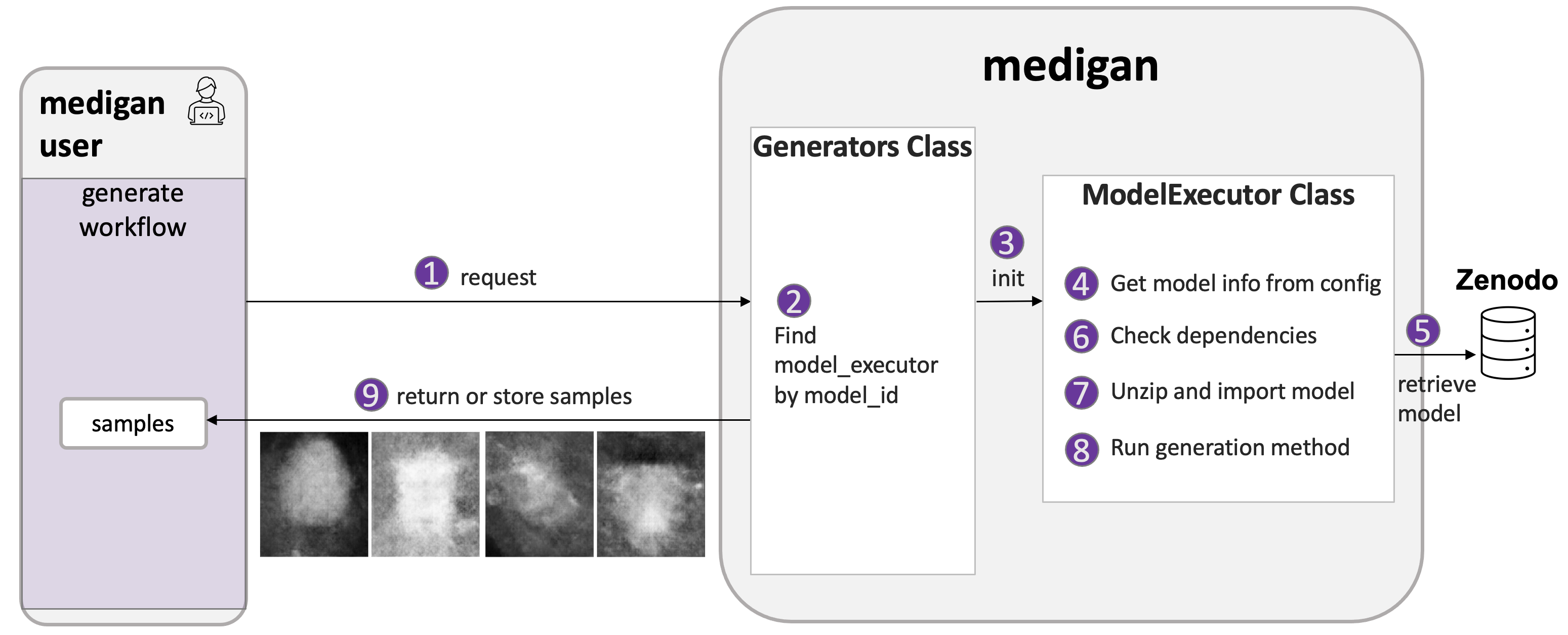}}
	 	 \caption{The generate workflow. A user specifies a \mintinline[fontsize=\footnotesize]{python}{model_id} in a request (1) to the \mintinline[fontsize=\footnotesize]{python}{Generators} class, which checks (2) if the model's \mintinline[fontsize=\footnotesize]{python}{ModelExecutor} class instance is already initialised. If not, a new one is created (3), which (4) gets the model's config from the \textit{global.json} dict, (5) loads the model (e.g., from \textit{Zenodo}), (6) checks its dependencies, and (7) unzips and imports it, before running its internal generate function (8). Lastly, the generated samples are returned to the user.
	 	 }
  	\label{fig:generate_workflow}
\end{figure}

An example of the usage of the \mintinline[fontsize=\footnotesize]{python}{generate} method is shown in Code Snippet \ref{code:generate}, which triggers the model execution workflow illustrated in Figure \ref{fig:generate_workflow}.
Further parameters of the generate method allow users to specify the number of samples to be generated (\mintinline[fontsize=\footnotesize]{python}{num_samples}), if samples are returned as list or stored on disk (\mintinline[fontsize=\footnotesize]{python}{save_images}), where they are stored (\mintinline[fontsize=\footnotesize]{python}{output_path}), and whether model dependencies are automatically installed (\mintinline[fontsize=\footnotesize]{python}{install_dependencies}). Optional model-specific inputs can be provided via the \mintinline[fontsize=\footnotesize]{python}{**kwargs} parameter. These include for example, (i) a non-default path to the model weights, (ii) a path to an input image \textcolor{mycorrect}{folder} for image-to-image translation models, (iii) a conditional input for class-conditional generative models, or (iv) the \mintinline[fontsize=\footnotesize]{python}{input_latent_vector} as commonly used as model input in GANs.

Running the \mintinline[fontsize=\footnotesize]{python}{generate} method triggers the \mintinline[fontsize=\footnotesize]{python}{generators} module to initialise a \mintinline[fontsize=\footnotesize]{python}{model_executor} instance for the user-specified generative model. The model is identified via its \mintinline[fontsize=\footnotesize]{python}{model_id} as unique key in the \textit{global.json} model metadata database, parsed and managed by the \mintinline[fontsize=\footnotesize]{python}{config_manager} module. Using the latter, the \mintinline[fontsize=\footnotesize]{python}{model_executor} checks if the required python package dependencies are installed, retrieves the Zenodo URL and downloads, unzips, and imports the model package. It further retrieves the name of the internal data generation function inside the model's \mintinline[fontsize=\footnotesize]{python}{__init}\mintinline[fontsize=\footnotesize]{python}{__.py} script. As final step before calling this function, its \textcolor{mycorrect}{parameters and} 
 their default values are retrieved from the metadata and combined with user-provided arguments. \textcolor{mycorrect}{These user-provided arguments customise the generation process, which enables handling of multiple image generation scenarios. For instance, the aforementioned provision of the input image folder allows users to point to their own images to transform them using \textit{medigan} models that are, for instance, pretrained for cross-modality translation.}
In the case of large dataset generation, the number of samples indicated by \mintinline[fontsize=\footnotesize]{python}{num_samples} are chunked into smaller sized batches and iteratively generated to avoid overloading the random-access memory available on the user's machine.

\begin{code}
\captionof{listing}{Executing a \textit{medigan} model for synthetic data generation.}
\vspace*{-5mm}
\label{code:generate}
\begin{minted}[mathescape,
               linenos,
               numbersep=5pt,
               gobble=0,
               frame=lines, 
               fontsize=\footnotesize,
               framesep=2mm]{python}
from medigan import Generators
generators = Generators()
# create 100 polyps with masks using model 10 (FASTGAN) 
generators.generate(model_id=10, num_samples=100)
\end{minted}
\end{code}

\subsubsection{Generate Workflow Extensions} \label{sec:generate_extensions}

Apart from storing or returning samples, a callable of the model's internal generate function  can be returned to the user by setting \mintinline[fontsize=\footnotesize]{python}{is_gen_function_returned}. This function with prepared but adjustable default arguments enables integration of the generate method into other workflows within \textit{medigan} (e.g., model visualisation) or outside of \textit{medigan} (e.g., a user's AI model training).
As further alternative, a torch\cite{pytorchPaszke} dataset or dataloader can be returned for any model in medigan running \mintinline[fontsize=\footnotesize]{python}{get_as_torch_dataset} or \mintinline[fontsize=\footnotesize]{python}{get_as_torch_dataloader}, respectively. This further increases the versatility with which users can introduce \textit{medigan}'s data synthesis capabilities into their AI model training and data preprocessing pipelines.

Instead of a user manually selecting a model via \mintinline[fontsize=\footnotesize]{python}{model_id}, a model can also be automatically selected based on the recommendation from the model search and/or ranking methods. For instance, as triggered by Code Snippet \ref{code:rank_search_generate}, the models found in a search for \textit{mammography} are ranked in ascending order based on \textit{FID}, with the highest ranking model being selected and executed to generate the synthetic dataset.

\begin{code}
\captionof{listing}{Sequential searching, ranking, and data generation with highest ranked model.}
\vspace*{-5mm}
\label{code:rank_search_generate}
\begin{minted}[mathescape,
               linenos,
               numbersep=5pt,
               gobble=0,
               frame=lines, 
               fontsize=\footnotesize,
               framesep=2mm]{python}
from medigan import Generators
generators = Generators()
values=['mammography'] # keywords for searching
metric='FID' # metric for ranking
generators.find_models_rank_and_generate(values=values, metric=metric)
\end{minted}
\end{code}

\subsection{Model Visualisation} \label{sec:visualisation}
\begin{figure}
	\centering
       	\frame{\includegraphics[width=0.9\textwidth]{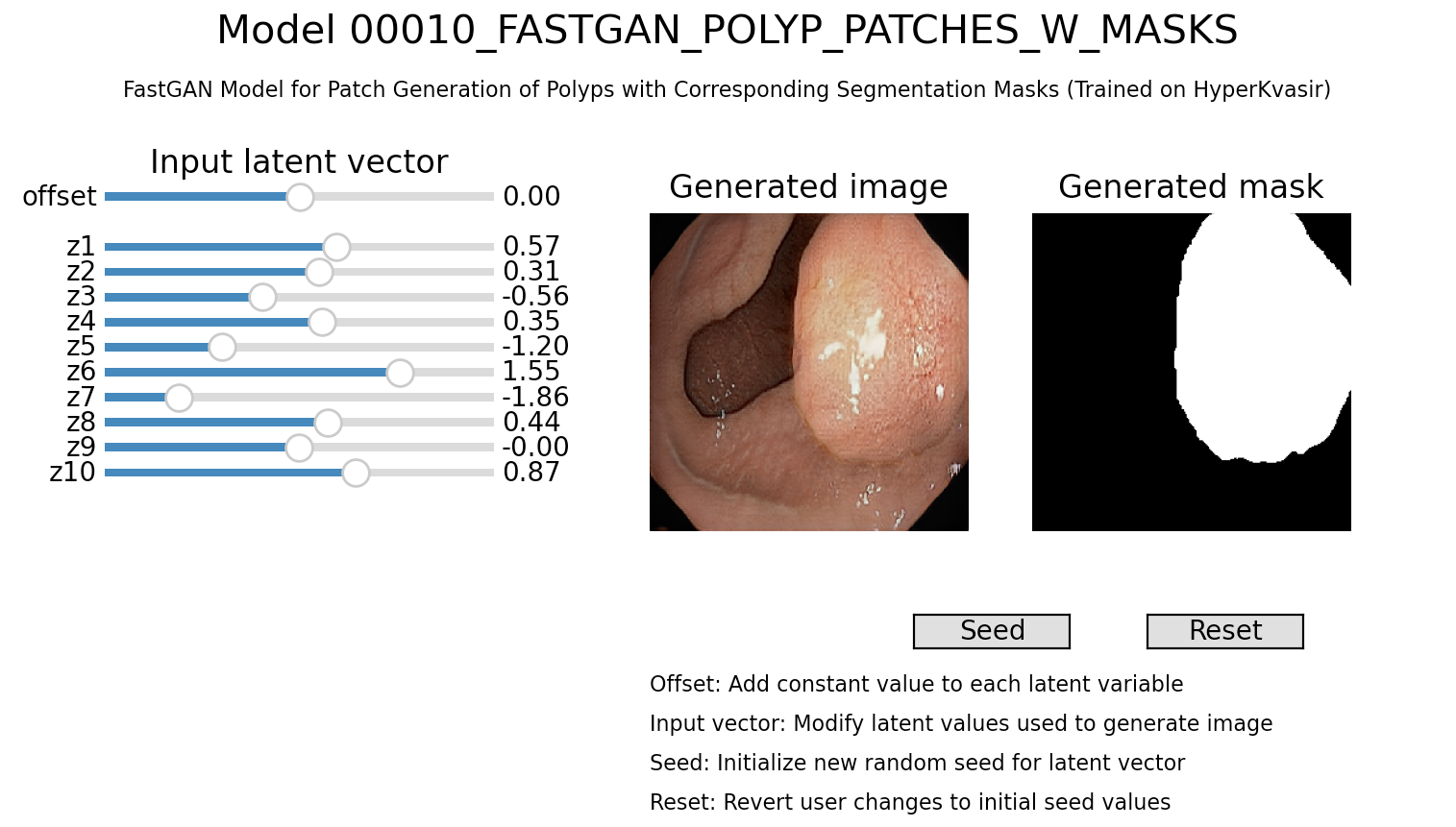}}
	 \caption{Graphical user interface of \textit{medigan}'s model visualisation tool on the example of model 10, a FastGAN that synthesises endoscopic polyp images with respective masks\cite{thambawita2022singan}. The latent input vector can be adjusted via the sliders, reset via the \textit{Reset} button and sampled randomly via the \textit{Seed} button.}
  	\label{fig:visualization_interface}
\end{figure}

To allow users to explore the generative models in \textit{medigan}, a novel model visualisation module has been integrated into the library. It allows to examine how changing inputs like the latent variable $z$ and/or the class conditional label $y$ (e.g. malignant / benign) affect the generation process.
Also, the correlation between multiple model outputs, such as the image and corresponding segmentation mask, can be observed and explored. Figure \ref{fig:visualization_interface} illustrates an example showing an image-mask sample pair from medigan's polyp generating FastGAN model\cite{thambawita2022singan}. This depiction of the graphical user interface (GUI) of the model visualisation tool can be recreated by running Code Snippet \ref{code:visualize}.

Internally, the \mintinline[fontsize=\footnotesize]{python}{model_visualizer} module retrieves a model's internal generate method as callable from the  \mintinline[fontsize=\footnotesize]{python}{model_executor} and adjusts the input parameters based on user interaction input from the GUI. This interaction further provides insight into a model's performance and capabilities. On one hand, it allows to assess the fidelity of the generated samples. On the other hand, it also shows the model's captured sample diversity, i.e., as observed output variation over all possible input latent vectors. We leave the automation of manual visual analysis of this output variation to future work. For instance, such future work can use the \mintinline[fontsize=\footnotesize]{python}{model_visualizer} to measure the variance of a reconstruction/perceptual error computed between pairs of images sampled from fixed-distance pairs of latent space vectors $z$.

The slider controls on the left of the interface allow to change the latent variable which for this specific model affects, for instance, polyp size, position, and background. As the size of the latent vector $z$ commonly is relatively large, each $n$ (e.g., 10) variables are grouped into one indexed slider resulting in $z_{m}$ adjustable latent input variables. The \textit{Seed} button on the right allows to initialise a new set of latent variables which results in a new generated image. The \textit{Reset} buttons allows to revert user's modifications to previous random values.

\begin{code}
\captionof{listing}{Visualisation of a model in \textit{medigan}.}
\vspace*{-5mm}
\label{code:visualize}
\begin{minted}[mathescape,
               linenos,
               numbersep=5pt,
               gobble=0,
               frame=lines, 
               fontsize=\footnotesize,
               framesep=2mm]{python}
from medigan import Generators
generators = Generators()
n = 10 # grouping latent vector z dimensions by dividing them by 10.
generators.visualize(model_id=10, slider_grouper=n) # polyp with mask
\end{minted}
\end{code}

\subsection{Model Contribution}
A core idea of \textit{medigan} is to provide a platform where researchers can share and access trained models via a standardised interface. 
We provide in-depth instructions 
on how to contribute a model to medigan complemented by implementations automating parts of the model contribution process for users.
In general, a pretrained model in medigan consists of a python \mintinline[fontsize=\footnotesize]{python}{__init}\mintinline[fontsize=\footnotesize]{python}{__.py} and, in case the generation process is based on a machine learning model, a respective checkpoint or weights file. The former needs to contain a synthetic data storage method and a data generation method with a set of standardised parameters described in Section \ref{sec:generate}. Ideally, a model package further contains a license file, a \textit{metadata.json} and/or a a \textit{requirements.txt} file, and a \textit{test.sh} script to quickly verify the model's functionalities. To facilitate creation of these files, \textit{medigan}'s GitHub repository provides model contributors with reusable templates 
for each of these files.

\begin{figure}
	\centering
       	 \scalebox{0.95}{\includegraphics[width=1.0\textwidth]{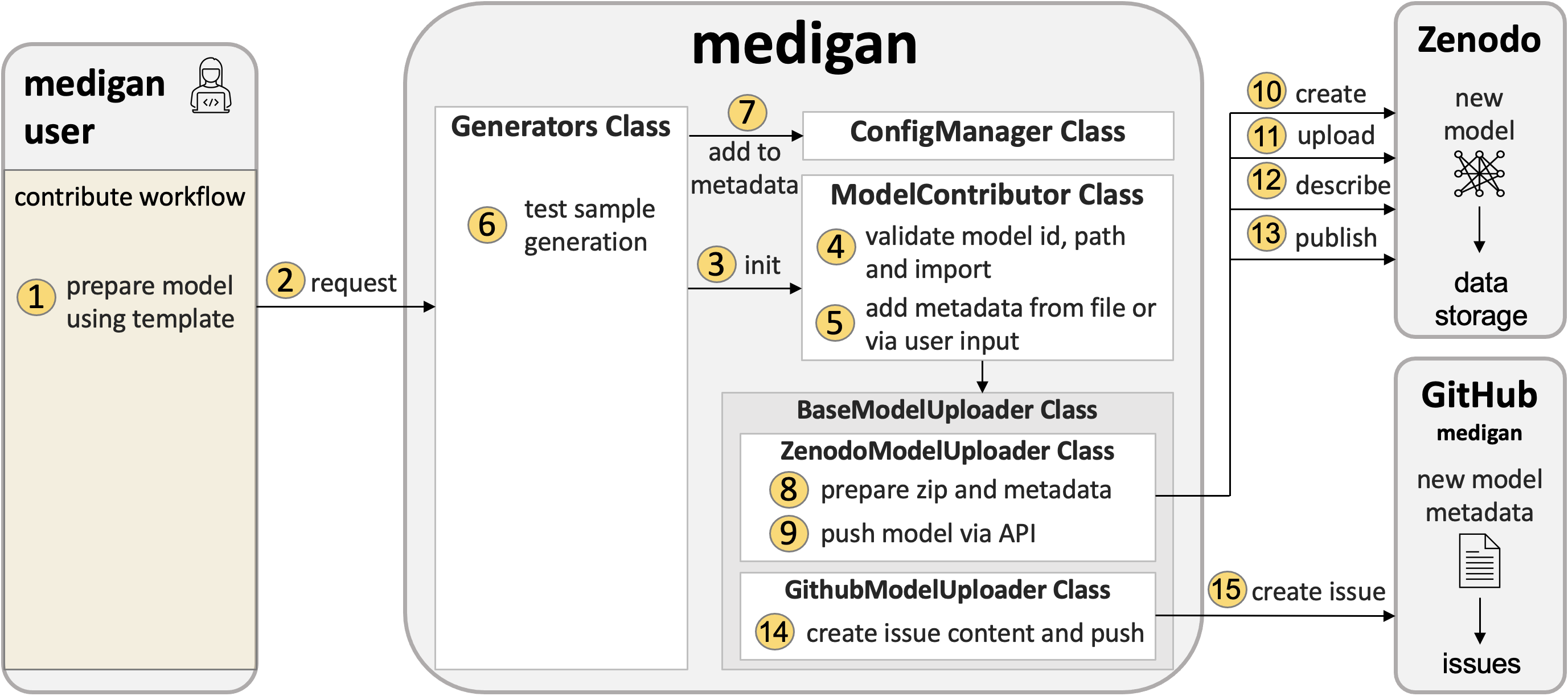}}
	 	 \caption{Model contribution workflow. After model preparation (1), a user provides the model's id and metadata (2) to the \mintinline[fontsize=\footnotesize]{python}{Generators} class to (3) initialise a \mintinline[fontsize=\footnotesize]{python}{ModelContributor} instance, which (4) validates and (5) extends the metadata. Next, (6) the model's sample generation capability is tested after (7) integration into \textit{medigan}'s \textit{global.json} model metadata. If successful, (8) the model package is prepared and (9-13) pushed to Zenodo via API. Lastly, (14-15) a GitHub issue containing the model metadata is created, assigned, and pushed to the \textit{medigan} repository.
	 	 }
  	\label{fig:contribute_workflow}
\end{figure}

Keeping the effort of pretrained model inclusion to a minimum, the \mintinline[fontsize=\footnotesize]{python}{generators} module contains a \mintinline[fontsize=\footnotesize]{python}{contribute} function that initialises a \mintinline[fontsize=\footnotesize]{python}{ModelContributor} class instance dedicated towards automating the remainder of the model contribution process. This includes automated 
(i) validation of the user-provided \mintinline[fontsize=\footnotesize]{python}{model_id},
(ii) validation of the path to the model's \mintinline[fontsize=\footnotesize]{python}{__init}\mintinline[fontsize=\footnotesize]{python}{__.py}, 
(iii) test of \mintinline[fontsize=\footnotesize]{python}{importlib} import of the model as package,
(iv) creation of the model's metadata dictionary,
(v) adding the model metadata to \textit{medigan}'s \textit{global.json} metadata,
(vi) end-to-end test of model with sample generation via \mintinline[fontsize=\footnotesize]{python}{generators.test_model()},
(vii) upload of zipped model package to Zenodo via API,
(viii) creation of a GitHub issue, which contains the Zenodo link and model metadata, in the \textit{medigan} repository. 
Being assigned to this GitHub issue, the \textit{medigan} development team is notified about the new model, which can then be added via pull request. Code Snippet \ref{code:contribute} shows how a user can run the \mintinline[fontsize=\footnotesize]{python}{contribute} method illustrated in Figure \ref{fig:contribute_workflow}. 
\begin{code}
\captionof{listing}{Contribution of a model to \textit{medigan}.}
\vspace*{-5mm}
\label{code:contribute}
\begin{minted}[mathescape,
               linenos,
               numbersep=5pt,
               gobble=0,
               frame=lines, 
               fontsize=\footnotesize,
               framesep=2mm]{python}
from medigan import Generators 
generators = Generators()
generators.contribute(
    model_id = "00100_YOUR_MODEL", # assign ID
    init_py_path ="path/ending/with/__init__.py", # model package root
    generate_method_name = "generate", # method inside __init__.py
    model_weights_name = "10000",
    model_weights_extension = ".pt",
    dependencies = ["numpy", "torch"],
    zenodo_access_token = "TOKEN", #zenodo.org/account/settings/applications
    github_access_token = "TOKEN") #github.com/settings/tokens
\end{minted}
\end{code}

\subsection{Model Testing Pipeline} \label{sec:pipeline}
Each new model contribution is being systematically tested before becoming part of \textit{medigan}. For instance, on each submitted pull request to medigan's GitHub repository, a CI pipeline automatically builds, formats, lints, and tests \textit{medigan}'s codebase. This includes the automatic verification of each model's package, dependencies, compatibility with the interface and correct functioning of its generation workflow. This allows to ensure that all models and their metadata in the \textit{global.json} file, are available and working in a reproducible and standardised manner.

\section{Applications}

\subsection{Community-Wide Data Access: Sharing the Essence of Restricted Datasets}

\begin{table}
\caption{Models currently available in \textit{medigan}. Also, computed FID scores for each model in \textit{medigan} are shown. 
The number of real samples used for FID calculation is indicated by \#imgs. The \textit{lower bound} $FID_{rr}$ is computed between a pair of randomly sampled sets of real data (real-real), while the \textit{model} $FID_{rs}$ is computed between two randomly sampled sets of real and synthetic data (real-syn). The results for model 7 (Flair, T1, T1c, T2) and 21 (T1, T2) are averaged across modalities.} 
\label{Table:model-table}
\centering
\scalebox{0.66}{
\begin{tabular}{clcccccccc} \toprule
&&&&&& \multicolumn{4}{c} {$FID_{ImageNet}$\cite{szegedy2016rethinking, deng2009imagenet}} \\
\cmidrule(r){7-10}
ID &  Output                       & Modality &     Model     &   Size   & Training dataset &  \#imgs & real-real & real-syn & $r_{FID}$ \\
\midrule
1 & breast calcifications        &   mammography &     DCGAN     &  128x128 &   INbreast\cite{Moreira2012} & 1000 & 33.61 & 67.60 & 0.497  \\

2 & breast masses                 &   mammography &     DCGAN\cite{alyafi2020dcgans}     &  128x128 &    OPTIMAM\cite{halling2020optimam} & 1000 & 28.85 & 80.51 & 0.358    \\ 

3 & high/low density breasts &   mammography &    CycleGAN \cite{garrucho2022high}   & 1332x800 &     BCDR\cite{Lopez2012} & 74 & 65.94 & 150.16  & 0.439 \\ 

4 & breast masses with masks       &   mammography &    pix2pix    &  256x256 &     BCDR\cite{Lopez2012} & 199 & 68.22 &  161.17 & 0.423  \\ 

5 & breast masses                 &   mammography &     DCGAN\cite{szafranowska2022sharing}     &  128x128 &     BCDR\cite{Lopez2012} & 199 & 68.22 & 180.04 & 0.379   \\ 

6 & breast masses                 &   mammography &    WGAN-GP\cite{szafranowska2022sharing}    &  128x128 &     BCDR\cite{Lopez2012} & 199 & 68.22 & 221.30 & 0.308  \\

7 & brain tumours with masks     &   cranial MRI 
&   Inpaint GAN \cite{kim2021synthesis}    &  256x256 &     BRATS 2018\cite{menze2014multimodal} & 1000 & 30.73 & 140.02  & 0.219  \\

8 & breast masses (mal/benign)    &   mammography &    C-DCGAN     &  128x128 &     CBIS-DDSM\cite{lee2017curated} & 379 & 37.56 & 137.75  & 0.272     \\

9 & polyps with masks             &   endoscopy  &    PGGAN\cite{thambawita2022singan}   &  256x256 &  HyperKvasir\cite{borgli2020hyperkvasir} & 1000  & 43.31 & 225.85  & 0.192  \\

10 & polyps with masks             &   endoscopy  &    FastGAN\cite{thambawita2022singan} &  256x256 &     HyperKvasir\cite{borgli2020hyperkvasir}& 1000  & 43.31 & 63.99 & 0.677    \\

11 & polyps with masks             &   endoscopy  &    SinGAN\cite{thambawita2022singan} &  \scriptsize{$\approx$}\normalsize{250x250} &     HyperKvasir\cite{borgli2020hyperkvasir} & 1000 &  43.31 & 171.15 & 0.253   \\

12 & breast masses (mal/benign)    &   mammography &    C-DCGAN     &  128x128 &     BCDR\cite{Lopez2012}  & 199  & 68.22 & 205.29 & 0.332    \\

13 & high/low density breasts MLO &   mammography &    CycleGAN \cite{garrucho2022high}   & 1332x800 &     OPTIMAM\cite{halling2020optimam} & 358  & 65.75 & 101.09 & 0.650   \\

14 & high/low density breasts CC  &   mammography &    CycleGAN \cite{garrucho2022high}   & 1332x800 &     OPTIMAM\cite{halling2020optimam} & 350  & 41.61 &  73.77  & 0.564   \\

15 & high/low density breasts MLO &   mammography &    CycleGAN \cite{garrucho2022high}  & 1332x800 &     CSAW\cite{dembrower2020multi} & 192 & 74.96 & 162.67  & 0.461   \\

16 & high/low density breasts CC  &   mammography &    CycleGAN \cite{garrucho2022high}   & 1332x800 &     CSAW\cite{dembrower2020multi} & 202 & 42.68 & 98.38 & 0.434     \\

17 & lung nodules  &   chest x-ray &    DCGAN   & 128x128 &     NODE21\cite{ecem_sogancioglu_2021_5548363} & 1476 & 24.34 & 126.78  & 0.192    \\ 

18 & lung nodules  &   chest x-ray &    WGAN-GP   & 128x128 &     NODE21\cite{ecem_sogancioglu_2021_5548363}  & 1476 & 24.34 & 211.47  & 0.115   \\ 

19 & full chest radiograph  &   chest x-ray &    PGGAN   & 1024x1024 &     ChestX-ray14\cite{wang2017chestx} & 1000 & 28.74 & 96.74  & 0.297    \\

20 & full chest radiograph  &   chest x-ray &    PGGAN \cite{segal2021evaluating}  & 1024x1024 &     ChestX-ray14\cite{wang2017chestx} & 1000 & 28.33 & 52.17 & 0.543 \\

21 & brain scans (T1/T2)  &  cranial MRI &    CycleGAN \cite{joshi2022nn}  & 224x192 &     CrossMoDA 2021\cite{dorent2022crossmoda} & 1000 & 24.41 & 59.49 & 0.410  \\

\bottomrule
\end{tabular}
}
\end{table}

\textit{medigan} facilitates sharing and reusing trained generative models with the medical research community. 
On one hand, this reduces the need for researchers to re-train their own similar generative models, which can reduce the extensive carbon footprint\cite{selvan2022carbon} of deep learning in medical imaging.
On the other hand, this provides a platform for researchers and data owners to share their dataset distribution without sharing the real data points of the dataset. Put differently, sharing generative models trained on (and instead of) patient datasets not only is beneficial as data curation step \cite{szafranowska2022sharing}, but also minimises the need to share images and personal data directly attributable to a patient. 
In particular, the latter can be quantifiably achieved when the generative model is trained using a differential privacy guarantee \cite{dwork2014algorithmic, osuala2022data} before being added to \textit{medigan}.
By reducing the barriers posed by data sharing restrictions and necessary patient privacy protection regulation, \textit{medigan} unlocks a new paradigm of medical data sharing via generative models.
This places \textit{medigan} at the centre towards solving the well-known issue of data scarcity\cite{bi2019artificial, osuala2022data} in medical imaging. 

Apart from that, \textit{medigan}'s generative model contributors benefit from an increased exposure, dissemination, and impact of their work, as their generative models become readily usable by other researchers. 
As Table \ref{Table:model-table} illustrates, to date, \textit{medigan} consists of 21 pretrained deep generative models contributed to the community. Among others, these include 2 conditional DCGAN models, 6 domain translation CycleGAN models and 1 mask-to-image pix2pix model. The training data comes from 10 different medical imaging datasets. Various of the models were trained on breast cancer datasets including INbreast \cite{Moreira2012}, OPTIMAM \cite{halling2020optimam}, BCDR \cite{Lopez2012}, CBIS-DDSM \cite{lee2017curated} and CSAW \cite{dembrower2020multi}. Models allow to generate samples of different pixel resolutions ranging from regions-of-interest patches of size 128x128 and 256x256 to full images of 1024x1024 and 1332x800 pixels.

\subsection{Investigating Synthetic Data Evaluation Methods} \label{sec:investigating_evaluation}
A further application of \textit{medigan} is testing the properties of medical synthetic data. For instance, evaluation metrics for generative models can be readily tested in \textit{medigan}'s multi-organ, multi-modality, multi-model synthetic data setting. 

Compared to generative modelling, synthetic data evaluation is a less explored research area \cite{osuala2022data}. In particular, in medical imaging the existing evaluation frameworks, such as the Fréchet Inception Distance (FID) \cite{heusel2017gans} or the Inception Score (IS) \cite{salimans2016improved}, are often limited in their applicability, as mentioned in Section \ref{sec:evaluation}. The models in \textit{medigan} allow to compare existing and new synthetic data evaluation metrics and their validation in the field of medical imaging. Multi-model synthetic data evaluation allows to measure the correlation and statistical significance between synthetic data evaluation metrics and downstream task performance metrics. This enables the assessment of clinical usefulness of generative models on one hand and of synthetic data evaluation metrics on the other hand. In that sense, the metric itself can be evaluated including its variations when measured under different settings, datasets, or preprocessing techniques.
    
\begin{figure*}[tb]
\begin{tikzpicture}[scale=1.0,transform shape]
\centering
  \pgfplotsset{
      scale only axis,
  }
  
\begin{axis}[enlargelimits=0.05, 
        xlabel=FID: real-synthetic,
        ylabel=FID: real-real,
        grid=both,
        scale only axis=true,
        width=0.9\textwidth, 
        height=4.5cm,
        xtick={0,20,...,300},
        ytick={0,10,...,100},
        xmin=50, 
        xmax=225,
        ymin=0,
        ymax=90,
      ]
    \addplot[
        scatter/classes={a={blue}, b={pink}, c={green}, d={orange}}, 
        scatter, mark=*, only marks, 
        scatter src=explicit symbolic,
        nodes near coords*={\Label},
        visualization depends on={value \thisrow{label} \as \Label}
    ] table [meta=class] {
        x y class label
        67.60 33.61 a \footnotesize{1} 
        80.51 28.85 a \footnotesize{2}  
        150.16 65.94 a \footnotesize{3} 
        161.17 68.22 a \tiny{4}  
        180.04 68.22 a \footnotesize{5}  
        221.30 68.22 a \footnotesize{6}
        140.02 30.73 b \footnotesize{7}
        137.75 37.56 a \footnotesize{8}  
        225.85 43.31 d \footnotesize{9} 
        63.99 43.31 d \footnotesize{10}  
        171.15 43.31 d \footnotesize{11} 
        205.29 68.22 a \footnotesize{12}  
        101.09 65.75 a \footnotesize{13}  
        73.16 34.17 a \footnotesize{14}  
        162.67 74.96 a \footnotesize{15}  
        98.378 42.68 a \footnotesize{16} 
        126.78 24.34 c \footnotesize{17}  
        211.47 24.34 c \footnotesize{18}  
        96.74 28.74 c \footnotesize{19}  
        52.17 28.33 c \footnotesize{20}
        59.49 24.41 b \footnotesize{21} 
    };
    \addplot [thick, smooth, red, dashed] table[y={create col/linear regression={y=Y}}]{
        X Y class label 
        67.60 33.61 a \footnotesize{1} 
        80.51 28.85 a \footnotesize{2}  
        150.16 65.94 a \footnotesize{3} 
        161.17 68.22 a \footnotesize{4}  
        180.04 68.22 a \footnotesize{5}  
        221.30 68.22 a \footnotesize{6}
        140.02 30.73 b \footnotesize{7}
        137.75 37.56 a \footnotesize{8}  
        225.85 43.31 d \footnotesize{9}
        63.99 43.31 d \footnotesize{10}
        171.15 43.31 d \footnotesize{11} 
        205.29 68.22 a \footnotesize{12}  
        101.09 65.75 a \footnotesize{13}  
        73.16 34.17 a \footnotesize{14}  
        162.67 74.96 a \footnotesize{15}  
        98.378 42.68 a \footnotesize{16} 
        126.78 24.34 c \footnotesize{17}  
        211.47 24.34 c \footnotesize{18}
        96.74 28.74 c \footnotesize{19} 
        52.17 28.33 c \footnotesize{20}
        59.49 24.41 b \footnotesize{21} 
    % ... 
    };
\end{axis}
\end{tikzpicture}
    \caption[]{Scatter plot illustrating the $FID_{rs}$ of \textit{medigan}'s models (real-synthetic) compared to the lower bound $FID_{rr}$ between two sets of the model's respective training dataset (real-real). The lower bound can represent an optimally achievable model and, as such, facilitates interpretation. Each model is represented by a dot below its model ID. The dots' colour encoding depicts model modality, where blue: Mammography, orange: Endoscopy, green: Chest x-ray, and pink: Brain MRI. The red regression line illustrates the trend across all data points/models.} 
    \label{fig:Scatter1}
\end{figure*}
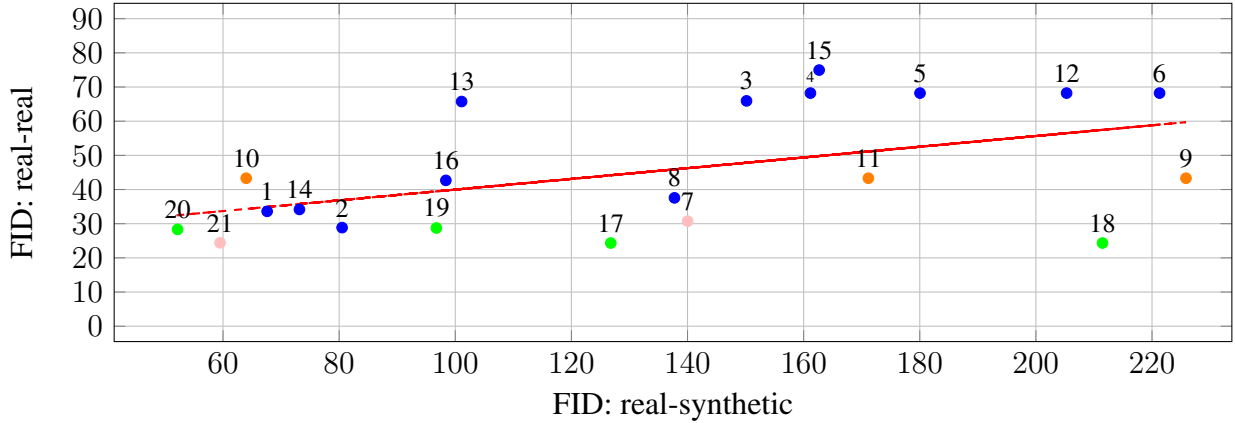

\subsubsection{FID of \textit{medigan} Models} \label{sec:evaluation_FID}
We compute the FID to assess the models in \textit{medigan} 
and report the results in Table \ref{Table:model-table}. We further note that the FID can be computed not only between a synthetic and a real dataset ($rs$), but also between two sets of samples of the real dataset ($rr$). As the $FID_{rr}$ describes the distance within two randomly sampled sets of the real data distribution, it can be used as an estimate of the real data variation and optimal lower bound for the $FID_{rs}$ as shown in Table \ref{Table:model-table}. Given the above, it follows that a high $FID_{rr}$ likely also results in a higher $FID_{rs}$, which highlights the importance of accounting for the $FID_{rr}$ when discussing the $FID_{rs}$. To do so, we propose the reporting of a FID ratio $r_{FID}$ to describe the $FID_{rs}$ in terms of the $FID_{rr}$.
    \begin{equation} \label{eq:5}
    \begin{aligned}
    r_{FID}(FID_{rs}, FID_{rr}) = 1- \frac{FID_{rs}-FID_{rr}}{FID_{rs}}, r_{FID} \in [0,1] \subset \mathbb{R}
    \end{aligned}
    \end{equation}
Assuming $FID_{rs} \geq FID_{rr}$ bounds $r_{FID}$ between 0 and 1, the $r_{FID}$ simplifies the comparison of FIDs computed using different models and datasets. A $r_{FID}$ close to 1 indicates that much of the $FID_{rs}$ can be explained by the general variation in the real dataset. The code used to compute the FID scores is available at  \href{https://github.com/RichardObi/medigan/blob/main/tests/fid.py}{https://github.com/RichardObi/medigan/blob/main/tests/fid.py}.

The models in Table \ref{Table:model-table} yielding the highest ImageNet-based $r_{FID}$ score are the ones with ID 10 (0.677, endoscopy, 256x256, FastGAN), ID 13 (0.650, mammography, 1332x800, CycleGAN), 14 (0.564, mammography, 1332x800, CycleGAN), 20 (0.543, chest x-ray, 1024x1024, PGGAN) and 1 (0.497, mammography, DCGAN, 128x128).
This indicates that the $r_{FID}$ does not depend on the modality, nor on the pixel resolution of the synthetic images. Further, neither image-to-image translation (e.g. CycleGAN) nor noise-to-image models (e.g. PGGAN, DCGAN, FastGAN) seem to have a particular advantage for achieving higher $r_{FID}$ results. 

The flow chart in Figure \ref{fig:Scatter1} provides further insight into the comparison between the lower bound $FID_{rr}$ and the model $FID_{rs}$. The red trend line shows a positive correlation between the $FID_{rr}$ and $FID_{rs}$, which corroborates our previous assumption that a higher model $FID_{rs}$ is to be expected given a higher lower bound $FID_{rr}$. Hence, for increased transparency, we motivate further studies to routinely report the lower bound $FID_{rr}$ and the FID ratio $r_{FID}$ apart from the model $FID_{rs}$. The 3-channel RGB endoscopic images represented by orange dots have a $FID_{rr}$ comparable to their grayscale radiologic counterparts. However, both chest x-ray datasets ChestX-ray14 \cite{wang2017chestx} and Node21 \cite{ecem_sogancioglu_2021_5548363} represented by green dots show a slightly lower $FID_{rr}$ than other modalities. The model $FID_{rs}$ shows a high variation across models without readily observable dependence on modality, generative model, or image size. 

\begin{table}
\caption{Normalised (left) and non-normalised (right) FID scores. This table measures the normalisation impact on FID scores based on a promising set of \textit{medigan}'s deep generative models. 
Synthetic samples were randomly-drawn for each model matching the number of available real samples.
The \textit{lower bound} $FID_{rr}$ is computed between a pair of randomly sampled sets of real data (real-real), while the \textit{model} $FID_{rs}$ is computed between two randomly sampled sets of real and synthetic data (real-syn). The results for model 7 (Flair, T1, T1c, T2) and 21 (T1, T2) are averaged across modalities.} 
\label{Table:FID_normalization}
\centering
\scalebox{0.69}{
\begin{tabular}{|| cc || cccccc || cccccc ||} \toprule
&&  \multicolumn{6}{c||}{\textbf{Normalised Images}} & \multicolumn{6}{c||}{\textbf{Non-Normalised Images}}  \\
&& \multicolumn{3}{c}  {$FID_{ImageNet}$\cite{szegedy2016rethinking, deng2009imagenet}} & \multicolumn{3}{c||}{$FID_{RadImageNet}$\cite{mei2022radimagenet}} & \multicolumn{3}{c} {$FID_{ImageNet}$\cite{szegedy2016rethinking, deng2009imagenet}} & \multicolumn{3}{c||}{$FID_{RadImageNet}$\cite{mei2022radimagenet}} \\
\cmidrule(r){3-5} \cmidrule(r){6-8} \cmidrule(r){9-11} \cmidrule(r){12-14}
ID 
& Dataset & real-real & real-syn & $r_{FID}$ & real-real &  real-syn & $r_{FID}$ & real-real & real-syn & $r_{FID}$ & real-real & real-syn & $r_{FID}$ \\ 
\midrule
1 
& Inbreast & 33.61 & 67.60 & 0.497 & 0.25 & 1.27 & 0.197
& 28.59 & 66.76 & 0.428 & 0.29 & 1.15 & 0.252 \\

2 
& Optimam & 28.85 & 80.51  & 0.358 & 0.22 & 6.19  & 0.036
& 28.75 & 77.95  & 0.369 & 0.33 & 4.11  & 0.080  \\

3 
& BCDR & 65.94 &  150.16  & 0.439  & 0.80 & 3.00  & 0.265 & 66.25 & 149.33  & 0.444 & 0.80 & 3.10  & 0.259 \\


5 
& BCDR & 68.22 & 180.04  & 0.379  & 0.99 & 1.67  & 0.593
& 64.45 & 174.38  & 0.370  & 0.87 & 4.04  & 0.215 \\

6 
& BCDR & 68.22 & 221.30  & 0.308  & 0.99 & 1.80  & 0.550
& 64.45 & 206.57  & 0.312 & 0.87 & 2.95  & 0.295  \\

7 
& BRATS 2018 & 30.73 & 140.02 & 0.219 & 0.07 & 5.31  & 0.012 & 
30.73  & 144.00 & 0.215  & 0.07 & 6.53 & 0.010 \\


8 
& CBIS-DDSM & 37.56 & 137.75  & 0.272 & 0.46 & 3.05  & 0.151
& 32.06 & 91.09  & 0.352 & 0.36 & 6.58 & 0.055  \\

10 
& HyperKvasir & 43.31 & 63.99 & 0.677 & 0.11 & 7.32  & 0.015
& 43.31 & 64.01 & 0.677 & 0.11 & 7.33 & 0.015  \\

12 
& BCDR & 68.22 & 205.29 & 0.332 & 0.99 & 5.69 & 0.080
& 64.45 & 199.50 & 0.323 & 0.87 & 4.25  & 0.205  \\

13 
& OPTIMAM & 65.75 &  101.01  & 0.650  & 0.17 & 1.14  & 0.153 & 65.83 & 101.15  & 0.651 & 0.18 & 1.10  & 0.163 \\

14 
& OPTIMAM & 41.61 & 73.77 & 0.564  & 0.16 & 0.83  & 0.190 & 41.71 & 74.03  & 0.563 & 0.15 & 0.81  & 0.184 \\

15 
& CSAW & 74.96 &  162.67  & 0.461  & 0.31 & 4.07  & 0.076 & 73.62 & 165.53  & 0.445 & 0.19 & 3.71  & 0.051 \\

16 
& CSAW & 42.68 &  98.38  & 0.439  & 0.38 & 2.71  & 0.142 & 42.50 & 99.81 & 0.426 & 0.22 & 2.82 & 0.077 \\

19 
& ChestX-ray14 & 28.75 & 96.74  & 0.297 & 0.19 & 0.77 & 0.243
& 28.75 & 96.78 & 0.297 & 0.19 & 0.66 & 0.286  \\

20 
& ChestX-ray14 & 28.33 & 52.17  & 0.543 & 0.20 & 2.83 & 0.071
& 28.33 & 52.38 & 0.541 & 0.20 & 2.59 & 0.077  \\

21 
& CrossMoDA & 24.41 & 59.49  & 0.410 & 0.02 & 1.45 & 0.014
& 24.41 & 60.11  & 0.406 & 0.02 & 1.40 & 0.014 \\


\bottomrule
\end{tabular}
}
\end{table}

\begin{figure*}[tb]
\begin{tikzpicture}[scale=1.0,transform shape]
\centering
  \pgfplotsset{
      scale only axis,
  }
\begin{axis}[enlargelimits=0.05, 
        xlabel=$FID_{ImageNet}$,
        ylabel=$FID_{RadImageNet}$,
        grid=both,
        scale only axis=true,
        width=0.9\textwidth, 
        height=4.5cm,
        xtick={0,10,...,300},
        ytick={0,1,...,10},
        xmin=50, 
        xmax=220,
        ymin=0,
        ymax=8.5,
      ]
    \addplot[
        scatter/classes={a={black}, b={cyan}},
        scatter, mark=*, only marks, sharp plot,
        scatter src=explicit symbolic,
        nodes near coords*={\Label},
        visualization depends on={value \thisrow{label} \as \Label}
    ] table [meta=class] {
        x y class label
        67.60 1.27 a \tiny{$1_{N}$} 
        66.76 1.15 b \footnotesize{} 
    };
        \addplot[
        scatter/classes={a={black}, b={cyan}},
        scatter, mark=*, only marks, sharp plot,
        scatter src=explicit symbolic,
        nodes near coords*={\Label},
        visualization depends on={value \thisrow{label} \as \Label}
    ] table [meta=class] {
        x y class label
        80.51 6.19 a \footnotesize{$2_{Norm}$}  
        77.95 4.11 b \footnotesize{} 
    };
    \addplot[
        scatter/classes={a={black}, b={cyan}},
        scatter, mark=*, only marks, sharp plot,
        scatter src=explicit symbolic,
        nodes near coords*={\Label},
        visualization depends on={value \thisrow{label} \as \Label}
    ] table [meta=class] {
        x y class label
        150.16 3.00 a \footnotesize{$3_{Norm}$}  
        149.33 3.10 b \footnotesize{} 
    };
    \addplot[
        scatter/classes={a={black}, b={cyan}},
        scatter, mark=*, only marks, sharp plot,
        scatter src=explicit symbolic,
        nodes near coords*={\Label},
        visualization depends on={value \thisrow{label} \as \Label}
    ] table [meta=class] {
        x y class label
        180.04 1.67 a \footnotesize{$5_{Norm}$}
        174.38 2.95 b \footnotesize{}
    };
    \addplot[
        scatter/classes={a={black}, b={cyan}},
        scatter, mark=*, only marks, sharp plot,
        scatter src=explicit symbolic,
        nodes near coords*={\Label},
        visualization depends on={value \thisrow{label} \as \Label}
    ] table [meta=class] {
        x y class label
        221.30 1.80 a \footnotesize{$6_{Norm}$}
        206.57 6.58 b \footnotesize{}
    };
    \addplot[
        scatter/classes={a={black}, b={cyan}},
        scatter, mark=*, only marks, sharp plot,
        scatter src=explicit symbolic,
        nodes near coords*={\Label},
        visualization depends on={value \thisrow{label} \as \Label}
    ] table [meta=class] {
        x y class label
        140.02 5.31 a \footnotesize{$7_{Norm}$}
        144.00  6.53 b \footnotesize{} 
    };
    \addplot[
        scatter/classes={a={black}, b={cyan}},
        scatter, mark=*, only marks, sharp plot,
        scatter src=explicit symbolic,
        nodes near coords*={\Label},
        visualization depends on={value \thisrow{label} \as \Label}
    ] table [meta=class] {
        x y class label
        137.75 3.05 a \footnotesize{$8_{Norm}$}  
        91.09 6.58 b \footnotesize{} 
    };
    \addplot[
        scatter/classes={a={black}, b={cyan}},
        scatter, mark=*, only marks, sharp plot,
        scatter src=explicit symbolic,
        nodes near coords*={\Label},
        visualization depends on={value \thisrow{label} \as \Label}
    ] table [meta=class] {
        x y class label
        63.99 7.32 a \footnotesize{$10_{Norm}$}  
        64.01 7.33 b \footnotesize{} 
    };
    \addplot[
        scatter/classes={a={black}, b={cyan}},
        scatter, mark=*, only marks, sharp plot,
        scatter src=explicit symbolic,
        nodes near coords*={\Label},
        visualization depends on={value \thisrow{label} \as \Label}
    ] table [meta=class] {
        x y class label
        205.29 5.69 a \footnotesize{$12_{Norm}$} 
        199.50 4.25 b \footnotesize{} 
    };
        \addplot[
        scatter/classes={a={black}, b={cyan}},
        scatter, mark=*, only marks, sharp plot,
        scatter src=explicit symbolic,
        nodes near coords*={\Label},
        visualization depends on={value \thisrow{label} \as \Label}
    ] table [meta=class] {
        x y class label
        101.09 1.14 a \footnotesize{$13_{Norm}$}  
        101.15 1.10 b \footnotesize{} 
    };
        \addplot[
        scatter/classes={a={black}, b={cyan}},
        scatter, mark=*, only marks, sharp plot,
        scatter src=explicit symbolic,
        nodes near coords*={\Label},
        visualization depends on={value \thisrow{label} \as \Label}
    ] table [meta=class] {
        x y class label
        73.79 0.83 a \tiny{$14_N$}  
        74.03 0.81 b \tiny{} 
    };
        \addplot[
        scatter/classes={a={black}, b={cyan}},
        scatter, mark=*, only marks, sharp plot,
        scatter src=explicit symbolic,
        nodes near coords*={\Label},
        visualization depends on={value \thisrow{label} \as \Label}
    ] table [meta=class] {
        x y class label
        162.67 4.07 a \footnotesize{$15_{Norm}$}  
        165.53 3.71 b \footnotesize{} 
    };
        \addplot[
        scatter/classes={a={black}, b={cyan}},
        scatter, mark=*, only marks, sharp plot,
        scatter src=explicit symbolic,
        nodes near coords*={\Label},
        visualization depends on={value \thisrow{label} \as \Label}
    ] table [meta=class] {
        x y class label
        98.38 2.71 a \footnotesize{$16_{Norm}$}  
        99.81 2.82 b \footnotesize{} 
    };
    \addplot[
        scatter/classes={a={black}, b={cyan}},
        scatter, mark=*, only marks, sharp plot,
        scatter src=explicit symbolic,
        nodes near coords*={\Label},
        visualization depends on={value \thisrow{label} \as \Label}
    ] table [meta=class] {
        x y class label
        96.75 0.77 a \tiny{$19_{N}$} 
        96.78 0.66 b \footnotesize{} 
    };
    \addplot[
        scatter/classes={a={black}, b={cyan}},
        scatter, mark=*, only marks, sharp plot,
        scatter src=explicit symbolic,
        nodes near coords*={\Label},
        visualization depends on={value \thisrow{label} \as \Label}
    ] table [meta=class] {
        x y class label
        52.17 2.83 a \footnotesize{$20_{Norm}$} 
        52.38 2.59 b \footnotesize{} 
    };
    \addplot[
        scatter/classes={a={black}, b={cyan}},
        scatter, mark=*, only marks, sharp plot,
        scatter src=explicit symbolic,
        nodes near coords*={\Label},
        visualization depends on={value \thisrow{label} \as \Label}
    ] table [meta=class] {
        x y class label
        59.49 1.45 a \tiny{$21_{N}$}
        60.11 1.40 b \footnotesize{} 
    };

    \addplot [thick, smooth, red, dashed] table[y={create col/linear regression={y=Y}}]{
       X Y class label
       67.60 1.27 a \footnotesize{$1_{Norm}$} 
       80.51 6.19 a \footnotesize{$2_{Norm}$}  
       150.16 3.00 a \footnotesize{$3_{Norm}$}  
       180.04 1.67 a \footnotesize{$5_{Norm}$}  
       221.30 1.80 a \footnotesize{$6_{Norm}$}
       140.02 5.31 a \footnotesize{$7_{Norm}$} 
       137.75 3.05 a \footnotesize{$8_{Norm}$}
       63.99 7.32 a \footnotesize{$10_{Norm}$}
       205.29 5.69 a \footnotesize{$12_{Norm}$}
       101.09 1.14 a \footnotesize{$13_{Norm}$}  
       73.79 0.83 a \footnotesize{$14_{Norm}$} 
       162.67 4.07 a \footnotesize{$15_{Norm}$}  
       98.38 2.71 a \footnotesize{$16_{Norm}$}
       96.75 0.77 a \tiny{$19_{N}$} 
       52.17 2.83 a \footnotesize{$20_{Norm}$}  
       59.49 1.45 a \footnotesize{$21_{Norm}$}
       66.76 1.15 a \footnotesize{$1_{Raw}$}
       77.95 4.11 a \footnotesize{$2_{Raw}$}
       149.33 3.10 b \footnotesize{3Raw} 
       174.38 2.95 a \footnotesize{$5_{Raw}$}  
       206.57 6.58 a \footnotesize{$6_{Raw}$}
       144.00  6.53 b \footnotesize{$7_{Raw}$} 
       91.09 6.58 a \footnotesize{$8_{Raw}$} 
       64.01 7.33 b \footnotesize{10Raw} 
       199.50 4.25 a \footnotesize{$12_{Raw}$}
       101.15 1.10 b \footnotesize{13Raw} 
       74.03 0.81 b \footnotesize{14Raw} 
       165.53 3.71 b \footnotesize{15Raw} 
       99.81 2.82 b \footnotesize{16Raw} 
       96.78 0.66 b \footnotesize{19Raw} 
       52.38 2.59 a \footnotesize{$20_{Raw}$}
       60.11 1.40 b \footnotesize{$21_{Raw}$} 
    };
       
\end{axis}
\end{tikzpicture}
    \caption[]{Scatter plot demonstrating the $FID_{rs}$ (real-synthetic) of \textit{medigan} models from Table \ref{Table:FID_normalization}.  The $FID_{rs}$ is based on the features of two different inception classifiers \cite{szegedy2016rethinking}, one trained on ImageNet \cite{deng2009imagenet} (x-axis) and the other trained on RadImageNet \cite{mei2022radimagenet} (y-axis). Each model is represented by a dot below its model ID. A black dot indicates a FID calculated from normalised (\textit{Norm/N}) images, e.g. with pixel values scaled between 0 and 1, as opposed to a blue dot indicating a FID calculated from images without previous normalisation. The dots that correspond to the same model IDs (normalised and non-normalised) are connected via black lines. The red regression line illustrates the trend across all data points.} 
    \label{fig:Scatter2}
\end{figure*}
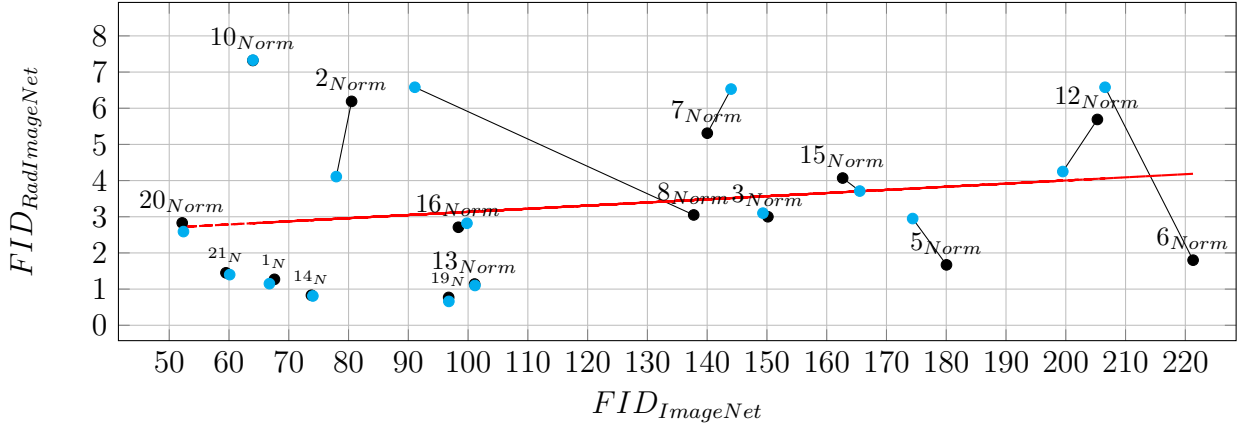

\subsubsection{Analysing Potential Sources of Bias in FID} \label{sec:evaluation_FID_bias}
The popular FID metric is computed based on the features of an Inception classifier (e.g., v1 \cite{szegedy2015going}, v3 \cite{szegedy2016rethinking}) trained on ImageNet \cite{deng2009imagenet} --- a database of natural images inherently different to the domain of medical images. This potentially limits the applicability of the FID to medical imaging data. Furthermore, the FID has been observed to vary based on the input image resizing methods and ImageNet backbone feature extraction model types \cite{kang2022studiogan}. Based on this, we further hypothesise a susceptibility of the FID to variation due to (a) different backbone feature extractor weights and random seed initialisations, (b) different medical and non-medical backbone model pretraining datasets, (c) different image normalisation procedures for real and synthetic dataset, (d) nuances between different frameworks and libraries used for FID calculation, and (f) the dataset sizes used to compute the FID.

Such variations can obstruct a reliable comparison of synthetic images generated by different generative models. Illustrating the potential of \textit{medigan} to analyse such variations, we report and experiment with the FID. In particular, we subject the FID to variations in (i) the pretraining dataset of its backbone feature extractor and by (ii) testing the effects of image normalisation across a set of \textit{medigan} models. We experiment with the Inception v3 model 
trained on the recent RadImageNet dataset \cite{mei2022radimagenet} released as radiology-specific alternative to the ImageNet database \cite{deng2009imagenet}. The RadImageNet-pretrained Inception v3 model weights we used are available at \href{https://github.com/BMEII-AI/RadImageNet}{https://github.com/BMEII-AI/RadImageNet}. We further compute the $FID_{rs}$ and $FID_{rr}$ with and without normalisation to analyse the respective impact on results. 

In Table \ref{Table:FID_normalization}, the FID results are summarised allowing for cross-analysis between variations due to image normalisation and/or due to the pretraining dataset of the FID feature extraction model. We observe generally lower FID values (1.15 to 7.32) for RadImageNet compared to ImageNet as FID model pretraining datasets (52.17 to 225.85). To increase FID comparability, we compute, as before, the FID ratio $r_{FID}$. The RadImageNet-based model results in notably lower $r_{FID}$ values for both normalised and non-normalised images. Notably, an exception to this are models with ID 5 (mammography, 128x128, DCGAN) and 6 (mammography, 128x128, WGAN-GP) achieving respective RadImageNet-based $r_{FID}$ scores of 0.593 and 0.550. In general, the RadImageNet-based model seems more robust at detecting if two sets of data originate from the same distribution resulting in low $FID_{rr}$ values. 
Overall, for most models, the FID is explained only by a limited amount by the variation in the real dataset and $r_{FID}<0.7$ for all ImageNet and RadImageNet-based FIDs. 

The scatter plot in Figure \ref{fig:Scatter2} further compares the RadImagnet-based FID with the ImageNet-FID for the models from Table \ref{Table:FID_normalization}. Noticeably, the difference between non-normalised and normalised images is surprisingly high for several models for both ImageNet and RadImageNet FIDs (e.g., models with IDs 6 and 8), while negligible for others (e.g., models with ID 1, 10, 13-16, and 19-21). Another observation is the relatively modest correlation between RadImageNet and ImageNet FID indicated by the slope of the red regression line. Counterexamples for this correlation include model 2 (normalised), which has a low ImageNet-based FID (80.51) compared to a high RadImageNet-based FID (6.19), and model 6 (normalised), which, in contrast, has a high ImageNet-based FID (221.30) and a low RadImageNet-based FID (1.80). With a low ImageNet-based FID (63.99), but surprisingly high RadImageNet-based FID (7.32), model 10 (both normalised and non-normalised) is a further counterexample. The example of model 10 is of particular interest, as it indicates limited applicability of the Radiology-specific RadImageNet-based FID for out-of-domain data, such as 3-channel endoscopic images.

Given the demonstrated high impact of backbone model training set and image normalisation on FID, it is to be recommended that studies specify the exact model used for FID calculation and any applied data preprocessing and normalisation steps. Further, where possible, reporting the RadImageNet-based FID allows for reporting a radiology domain-specific FID. The latter is seemingly less susceptible to variation in the real datasets than the ImageNet-based FID, while also being capable of capturing other, potentially complementary, patterns in the data.

\subsection{Improving Clinical Medical Image Analysis} \label{sec:improving_downstream}

\begin{table}
\caption{Examples of the impact of synthetic data generated by \textit{medigan} models on downstream task performance. Based on real test data, we compare the performance metrics of a model trained \textit{only on real} data with a model trained on real data \textit{augmented with synthetic} data. The metrics are taken from the respective publications describing the models.}
\label{Table:downstream_task1}
\centering
\scalebox{0.8}{
\begin{tabular}{ cccc  c  c } \toprule
ID 
& Test Dataset & Task & Metric & Trained on Real & Real + Synthetic  \\ 
\midrule
\midrule
2 
& OPTIMAM & Mammogram Patch Classification\cite{alyafi2020dcgans} & F1 & 0.90  & 0.96  \\
\midrule
3 
& BCDR & Mammogram Mass Detection\cite{garrucho2022high} & FROC AUC & 0.83  & 0.89  \\
\midrule

5 
& BCDR & Mammogram Patch Classification\cite{szafranowska2022sharing} & F1 & 0.891  & 0.920  \\
\midrule
5 
& BCDR & Mammogram Patch Classification\cite{szafranowska2022sharing} & AUROC & 0.928  & 0.959  \\
\midrule
5 
& BCDR & Mammogram Patch Classification\cite{szafranowska2022sharing} & AUPRC & 0.986  & 0.992  \\
\midrule

6 
& BCDR & Mammogram Patch Classification\cite{szafranowska2022sharing} & F1 & 0.891  & 0.969  \\
\midrule
6 
& BCDR & Mammogram Patch Classification\cite{szafranowska2022sharing} & AUROC & 0.928  & 0.978  \\
\midrule
6 
& BCDR & Mammogram Patch Classification\cite{szafranowska2022sharing} & AUPRC & 0.986  & 0.996  \\
\midrule
7 
& BRATS 2018 & Brain Tumour Segmentation\cite{kim2021synthesis} & Dice & 0.796 & 
0.814 \\ 
\midrule

14 
& OPTIMAM & Mammogram Mass Detection\cite{garrucho2022high} & FROC AUC & 0.83  & 0.85  \\
\midrule
15 
& OPTIMAM & Mammogram Mass Detection\cite{garrucho2022high}  & FROC AUC & 0.83  & 0.85  \\

\bottomrule
\end{tabular}
}
\end{table}

A high impact clinical application of synthetic data is the improvement of clinical downstream task performance such as classification, detection, segmentation, or treatment response estimation. This can be achieved by using image synthesis for data augmentation, domain adaptation and data curation (e.g. artifact removal, noise reduction, super-resolution) \cite{osuala2022data, diaz2021data} to enhance the performance of clinical decision support systems such as computer-aided diagnosis (CADx) and detection (CADe) software.

In Table \ref{Table:downstream_task1} the capability of improving the clinical downstream task performance is demonstrated for various \textit{medigan} models and modalities. Downstream task models trained on a combination of real and synthetic imaging data achieve promising results surpassing the alternative results achieved from training only on real data. The results are taken from the respective publications \cite{alyafi2020dcgans, garrucho2022high, szafranowska2022sharing, kim2021synthesis} and indicate that image synthesis can further improve the promising performance demonstrated by deep learning-based CADx and CADe systems\textcolor{mycorrect}{, e.g.,} in mammography \cite{Abdelrahman2021} and brain MRI \cite{menze2014multimodal}. 
\textcolor{mycorrect}{For downstream task evaluation, we generally note the importance of avoiding data leakage between training, validation and test sets by training the generative model either on only the dataset partition used to train the respective downstream task model (e.g., IDs 2, 3, 7, 14, 15) or to train the generative models on an entirely different dataset (e.g., IDs 5, 6).}

The approaches displayed in Table \ref{Table:downstream_task2} represent the application, where synthetic data is used instead of real data to train downstream task models. Despite an observable performance decrease when training on synthetic data only, the results \cite{thambawita2022singan, segal2021evaluating, joshi2022nn} demonstrate the usefulness of synthetic data if none or only limited real training data is available or shareable. For example, if labels or annotations in a target domain are scarce but present in a source domain, a generative model can translate annotated data from the source domain to the target domain to enable supervised training of downstream task models \cite{joshi2022nn, dorent2022crossmoda}.

\begin{table}
\caption{Examples of the impact of synthetic data generated by \textit{medigan} models on downstream task performance. Based on real test data, we compare the performance metrics of a model trained \textit{only on real} data with a model trained \textit{only on synthetic} data. The metrics are taken from the respective publications describing the models. \textit{n.a.} refers to the case where only synthetic data can be used, as no annotated real training data is available 
.}
\label{Table:downstream_task2}
\centering
\scalebox{0.8}{
\begin{tabular}{ cccc c c } \toprule
ID 
& Test Dataset & Task & Metric & Trained on Real & Trained on Synthetic \\ 
\midrule
4 
& BCDR & Mammogram Mass Segmentation  & Dice & 0.865 &
0.737 \\
\midrule
11 
& HyperKvasir & Polyp Segmentation\cite{thambawita2022singan} & Dice Loss & 0.112 & 0.137  \\
\midrule
11 
& HyperKvasir & Polyp Segmentation\cite{thambawita2022singan} & IoU & 0.827 & 0.798 \\
\midrule
11 
& HyperKvasir & Polyp Segmentation\cite{thambawita2022singan} & F-Score & 0.888 & 0.863  \\
\midrule
20 
& ChestX-ray14 & Lung Disease Classification\cite{segal2021evaluating} & AUROC &   0.947 & 0.878  \\
\midrule
21 
& CrossMoDA & Brain Tumour Segmentation\cite{joshi2022nn} & Dice & n.a.  & 0.712 \\ 
\midrule
21 
& CrossMoDA & Cochlea Segmentation\cite{joshi2022nn} & Dice & n.a.  & 0.478 \\ 

\bottomrule
\end{tabular}
}
\end{table}

\section{Discussion and Future Work}

In this work, we introduced \textit{medigan}, an open-source Python library, 
which allows to share pretrained generative models for synthetic medical image generation. The package is easily integrable into other packages and tools, including commercial ones. Synthetic data can enhance the performance, capabilities, and robustness of data-hungry deep learning models as well as to mitigate common issues such as domain shift, data scarcity, class imbalance, and data privacy restrictions. Training one's own generative network can be complex and expensive since it requires a considerable amount of time, effort, specific dedicated hardware, carbon emissions, as well as knowledge and applied skills in generative AI. An alternative and complementary solution is the distribution of pretrained generative models to allow their reuse by AI researchers and engineers worldwide.

\textit{medigan} can help to reduce the time to run synthetic data experiments and can readily be added as a component, e.g., as a dataloader as discussed in Section \ref{sec:generate_extensions}, in AI training pipelines. As such, the generated data can be used to improve supervised learning models as described in Section \ref{sec:improving_downstream} via training or fine-tuning, but can also serve as plug-and-play data source for self/semi-supervised learning, e.g., to pretrain clinical downstream task models.

Furthermore, studies that use additional synthetic training data for training deep learning models often not report all the specifics about their underlying generative model \cite{osuala2022data, lekadir2021futureai}. Within \textit{medigan}, each generative model is documented, openly accessible and re-usable. This increases the reproducibility of studies that use synthetic data and make it more transparent where the data or parts thereof originated from. This can help to achieve the traceability objectives outlined in the FUTURE-AI consensus guiding principles towards AI trustworthiness in medical imaging \cite{lekadir2021futureai}. \textit{medigan}'s currently 21 generative models are illustrated in Table \ref{Table:model-table} and developed and validated by AI researchers and/or specialised medical doctors. Furthermore, each model contains traceable \cite{lekadir2021futureai} and version-controlled metadata in medigan's \textit{global.json} file, as outlined in Section \ref{sec:metadata}. The searchable (see Section \ref{sec:search_workflow}) metadata allows to choose a suitable model for a user's task at hand and includes, among others, the dataset used during the training process, the trained date, publication, modality, input arguments, model types, and comparable evaluation metrics.

\textcolor{mycorrect}{To assess model suitability, users are recommended to first (i) ensure the compatibility between their planned downstream task (e.g., mammogram region-of-interest classification) and a candidate medigan model (e.g., mammogram region-of-interest generator). Secondly, (ii) a user's real (test) data and the model's synthetic data should be compatible corresponding, for instance, in domain, organ, or disease manifestation. If the awareness of the domain shifts between real and synthetic data remains limited after this qualitative analysis, (iii) a quantitative assessment (e.g., via FID) is recommended. Finally, (iv) it is to be assessed if a downstream task improvement is plausible. This depends, among others, on the tested scenario and the task at hand, but also on the amount, domain, task specificity and quality of the available real data, and the generative model's capabilities as indicated by its reported evaluation metrics from previous studies. If a positive impact of synthetic data on downstream task performance is plausible, users are recommended to proceed towards empirical verification.}

The exploration and multi-model evaluation of the properties of generative models and synthetic data is a further application of \textit{medigan}. 
\textit{medigan}'s visualisation tool (see Section \ref{sec:visualisation}) intuitively allows user to explore and adjust the input latent vector of generative models to visually evaluate, for instance, its inherent diversity and condition adherence\cite{osuala2022data} (i.e. how well does a given mask or label fit to the generated image).
The evaluation of synthetic data by human experts, such as radiologists, is a costly and time-consuming task, which motivates the usage of automated metric-based evaluation such as the FID. Our multi-model analysis reveals sources of bias in FID reporting. We show the susceptibility of FID to vary substantially based on changes in input image normalisation or in the choice of the pretraining dataset of the FID feature extractor. This finding highlights the need to report the specific models, pre-processing and implementations used to compute the FID alongside the FID ratio $r_{FID}$ proposed in Section \ref{sec:evaluation_FID} to account for the variation immanent in the real dataset. With \textit{medigan} model experiments demonstrably leading to insights in synthetic data evaluation, future research can use \textit{medigan} as a tool to accelerate, test, analyse, and compare new synthetic data and generative model evaluation and exploration techniques.

\subsection{Legal Frameworks for Sharing of Synthetic and Real Patient Data}

Many countries have enacted regulations that govern the use and sharing of data related to individuals. The two most recognised legal frameworks are the Health Insurance Portability and Accountability Act (HIPAA) \cite{hipaa} from the United States (U.S.) and the General Data Protection Regulation (GDPR) \cite{gdpr2018reg} from the European Union (E.U.). These regulations govern the use and disclosure of individuals' protected health information (PHI) and assures individuals' data is protected while allowing use for providing quality patient care \cite{nass2009beyond,usdhhs2003summary,shah2020secondary,kubben2019gdpr}.

Conceptually, synthetic data is not real data about any particular individual and conversely to real data, synthetic data can be generated at high volumes and potentially shared without restriction. In this sense, under both GDPR and HIPAA regulation, the rules govern the use of real data for the generation and evaluation of synthetic datasets, as well as the sharing of the original dataset. However, once fully synthetic data is generated, this new dataset falls outside the scope of the current regulations based on the argument that there is no direct correlation between the original subjects and the synthetic subjects. A common interpretation is that as long as the real data remains in a secure environment during the generation of synthetic data, there is little to no risk to the original subjects \cite{elemam2020practical}.

As a consequence, the use of synthetic data can help prevent researchers from inadvertently using and possibly exposing patients identifiable data. Synthetic data can also lessen the controls imposed by Institutional Review Boards (IRBs) and based on international regulations by ensuring data is never mapped to real individuals \cite{dankar2021fake}. There are multiple methods of generating synthetic data, some of which include building models from real data which can create a set statistically similar to real data. How similar the synthetic data is to real word data often defines its "utility", which will vary depending on the synthesis methods used and the needs of the study at hand. If the utility of the synthetic data is high enough then evaluation results are expected to be similar to the those that use real data \cite{elemam2020practical}. Being built based on real data, a common concern is patient re-identification and leaking of patient-specific features in generative models \cite{stadler2022synthetic, osuala2022data}. Despite the arguably permissive aforementioned regulations, de-identification \cite{diaz2021data} of the training data prior to generative model training is to be recommended. This can minimise the possibility of generative models leaking sensitive patient data during inference and after sharing. A further recommended and mathematically-proven tool for privacy preservation is differential privacy (DP) \cite{dwork2014algorithmic}. DP can be included in the training of deep generative model, among other setups, by adding DP noise to the gradients.

\subsection{Expansion of Available Models}
In the future, further generative models across medical imaging disciplines, modalities and organs can be integrated into medigan. The capabilities of additional models can range from privatising or translating the user's data from one domain to another, balancing or de-biasing imbalanced datasets, reconstructing, denoising or removing artifacts in medical images, \textcolor{mycorrect}{or resizing images e.g. using image super-resolution techniques}. Despite \textit{medigan}'s current focus on models based on Generative Adversarial Networks \cite{goodfellow2014generative}, the inclusion of different additional types of generative models is desirable and will enable insightful comparisons. In particular, this is to be further emphasised considering the recent successes of Diffusion Models \cite{sohl2015deep, song2019generative, ho2020denoising}, Variational Autoencoders \cite{kingma2013auto}, and Normalizing Flows \cite{rezende2015variational, dinh2014nice, dinh2016density} in the computer vision and medical imaging \cite{pinaya2022brain, pinaya2021unsupervised, pawlowski2020deep} domains. \textcolor{mycorrect}{Before integrating and testing a new model via the pipeline described in \ref{sec:pipeline}, we assess whether a model is to become a candidate for inclusion into medigan. This three-fold assessment is based on the SynTRUST framework\cite{osuala2022data} and reviews whether (1) the model is well-documented (e.g. in a respective publication), (2) the model or its synthetic data is applicable to a task of clinical relevance, and (3) whether the model has been methodically validated.}

\subsection{Synthetic DICOM Generation}

Since the dominant data format used for medical imaging is DICOM (Digital Imaging and Communications in Medicine), we plan to enhance medigan by integrating the generation of DICOM compliant files. DICOM consists of two main components, pixel data and the DICOM header. The latter can be described as an embedded dataset rich with information related to the pixel data such as the image sequence, patient, physicians, institutions, treatments, observations, and equipment \cite{diaz2021data}. Future work will explore combining our GAN generated images with synthetic DICOM headers. The latter will be created from the same training images from which the \textit{medigan} models are trained to create synthetic DICOM data with high statistical similarity to real world data. In this regard, a key research focus will be the creation of an appropriate and DICOM-compliant description of the image acquisition protocol for a synthetic image. The design and development of an open-source software package for generating DICOM files based on synthesised DICOM headers associated to (synthetic) images will extend prior work \cite{rutherford2021dicom} that demonstrated the generation of synthetic headers for the purpose of evaluating de-identification methods. 

\section{Conclusion}


We presented the open-source \textit{medigan} package, which helps researches in medical imaging to rapidly create synthetic datasets for a multitude of purposes such as AI model training and benchmarking, data augmentation, domain adaptation, and inter-centre data sharing. \textit{medigan} provides simple functions and interfaces for users allowing to automate generative model search, ranking, synthetic data generation, and model contribution. By reuse and dissemination of existing generative models in the medical imaging community, \textit{medigan} allows researchers to speed up their experiments with synthetic data in a reproducible and transparent manner.

We discuss 3 key applications of \textit{medigan}, which include (i) sharing of restricted datasets, (ii) improving clinical downstream task performance, and (iii) analysing the properties of generative models, synthetic data, and associated evaluation metrics.
Ultimately, the aim of \textit{medigan} is to contribute to benefiting patients and clinicians, e.g., by increasing the performance and robustness of AI models in clinical decision support systems.

\subsection*{Disclosures}
The authors have no conflicts of interest to declare that are relevant to the content of this article.

\subsection* {Acknowledgements}
We would like to thank all model contributors, such as Alyafi et al (2020) \cite{alyafi2020dcgans}, Szafranowska et al (2022) \cite{szafranowska2022sharing}, Thambawita et al (2022) \cite{thambawita2022singan}, Kim et al (2021) \cite{kim2021synthesis}, Segal et al (2021) \cite{segal2021evaluating}, Joshi et al (2022) \cite{joshi2022nn}, and Garrucho et al (2022) \cite{garrucho2022high}. This project has received funding from the European Union’s Horizon 2020 research and innovation programme under grant agreement No 952103 and No 101057699. Eloy García and Kaisar Kushibar hold the Juan de la Cierva fellowship from the Ministry of Science and Innovation of Spain with reference numbers FJC2019-040039-I and FJC2021-047659-I, respectively.

\subsection* {Data, Materials, and Code Availability} 

\textit{medigan} is a free Python (v3.6+) package published under the MIT license and distributed via the Python Package Index (\href{https://pypi.org/project/medigan/}{https://pypi.org/project/medigan/}). The package is open-source and invites the community to contribute on GitHub (\href{https://github.com/RichardObi/medigan}{https://github.com/RichardObi/medigan}).
A detailed documentation of \textit{medigan} is available (\href{https://medigan.readthedocs.io/en/latest/}{https://medigan.readthedocs.io/en/latest/})
that contains installation instructions, the API reference, a general description, code examples, a testing guide, a model contribution user guide, and documentation of the generative models available in medigan.


\bibliography{report}   
\bibliographystyle{spiejour}   



\vspace{1ex}
\noindent Biographies and photographs of the other authors are not available.

\listoffigures
\listoftables

\end{spacing}

\end{document}